\documentclass[preprint,12pt]{elsarticle}

\usepackage{amsmath,amssymb}
\usepackage{lineno}
\usepackage{graphicx}
\usepackage{tabularx}
\usepackage{booktabs}
\usepackage{microtype}
\usepackage[colorlinks=true, linkcolor=blue!80!black, citecolor=blue!80!black, urlcolor=blue!80!black]{hyperref}
\usepackage{xurl}

\begin{document}

\begin{frontmatter}

\title{Developing a Subdistrict-Level Economic Progress Index Using Geospatial Data}

\author[1]{Meenu Mohan\corref{cor1}}
\ead{meenu.mohan@sattva.co.in}

\author[2]{Siddhartha Chattopadhyay}


\cortext[cor1]{Corresponding author.}

\address[1]{Sattva Media and Consulting, Domlur, Bengaluru, Karnataka, India}
\address[2]{Indian Institute of Technology Kharagpur, Kharagpur, India}

\begin{abstract}
\setlength{\parindent}{0pt}
\setlength{\parskip}{0.8em}
This paper develops an Economic Progress Index (EPI), a composite measure built from Earth Observation (EO) data with unsupervised Principal Component Analysis (PCA) weighting. The index captures the spatially observable productive and enabling capacity of local economies alongside the intensity with which economic space is used. It is designed for hyper-local sub-national monitoring: it tracks local economic activity and growth, and reveals spatial disparities and regional dynamics that national and state accounts often mask. In developing economies, sub-national monitoring is held back by sparse data. Where official accounts exist, Gross State and District Domestic Product (GSDP/GDDP) estimates suffer from multi-year survey lags and structural distortions, notably price-deflator noise, rigid service-sector productivity assumptions and top-down undercounting of the informal economy. These bottlenecks are sharper still at the subdistrict (taluka/tehsil) scale, where place-based policy needs high-frequency evidence.

The index fuses multi-sensor satellite data, gridded population and crowdsourced road and amenity networks across 357 SHRUG v2.1 subdistricts of Maharashtra (derived from 2011 Census), India, over 46 quarters (2015 Q1--2026 Q2). Seasonal-Trend decomposition using LOESS (STL) separates long-run capital accumulation from monsoon cloud artefacts and crop phenology. A non-negative PCA (NN-PCA) produces rankings nearly identical to standard PCA ($\rho = 0.998$) while giving sign-consistent, interpretable weights. Validated across 35 districts, the index explains 74\% of the variation in the NFHS-5 Wealth Index, 51\% in household consumption (HCES 2023--24) and over 78\% in real GDDP (all $p < 0.001$).

Because the index measures structural capacity rather than cyclical output, it does not depend on price deflators and complements GDP rather than copying it, and kept rising through the FY2020--21 COVID-19 contraction. Across Maharashtra's six revenue divisions, relative inequality fell by 44\% (CV 0.64 $\to$ 0.36) even as the absolute gap between the leading and lagging divisions widened by 32\%, and district-level $\beta$-convergence implies a half-life of about 14 years. The framework offers a scalable, price-invariant basis for structural monitoring and district-level nowcasting in data-sparse economies, with potential applications in infrastructure evaluation and municipal credit assessment. Future work will focus on scaling the index into a nationwide subdistrict-level monitoring framework with quarterly updates.
\end{abstract}

\begin{keyword}
Earth Observation \sep night-time lights \sep subdistrict economic activity \sep composite index \sep non-negative PCA \sep STL decomposition \sep India \sep regional convergence
\JEL R11 \sep R12 \sep O18 \sep E01 \sep C43 \sep C38
\end{keyword}

\end{frontmatter}

\section{Introduction}
\label{sec:introduction}

\subsection{Data Gaps in Sub-National Monitoring}
Reliable and timely socio-economic data are fundamental to evidence-based policymaking, resource allocation, and development monitoring. In India, large-scale surveys and administrative data systems provide valuable information on economic and social conditions. However, they are often constrained by infrequent update cycles, reporting lags, and limited spatial granularity. These limitations become particularly pronounced at the subdistrict level, where policy interventions are increasingly targeted but where high-frequency, comparable, and geographically consistent data remain scarce. Although initiatives such as the Aspirational Districts/Blocks Programme have strengthened localized monitoring and governance, they continue to rely predominantly on administrative indicators whose quality, coverage, and timeliness may vary across regions.

\subsection{Structural Limitations of Official District Accounts}
Critically, even where formal sub-national accounts exist, official Gross State and District Domestic Product (GSDP/GDDP) metrics are undermined by structural distortions. For instance, assuming uniform statewide service-sector productivity artificially inflates output figures in poorer districts where actual returns are lower \cite{Hatekar2024}. Furthermore, GSDP calculations remain vulnerable to ``head-office distortion''; attributing manufacturing value to corporate headquarters rather than actual operational sites, while struggling with outdated base years, deflator inaccuracies, and severe data gaps in the unorganized sector \cite{Sengupta2019b}.

While these constraints do not invalidate the utility of formal sub-national accounts, they underscore the necessity of integrating complementary frameworks which offer high-resolution, high-frequency visibility into localized economic trajectories, bridging the critical data gaps inherent in traditional administrative reporting.

\subsection{Earth Observation as a Spatial Proxy}
Earth Observation (EO), remote sensing, and geospatial analytics approaches offer new opportunities to complement traditional data systems. A growing body of research has demonstrated the potential of satellite-derived indicators to serve as proxies for economic activity, infrastructure development, and environmental factors. Unlike conventional survey-based measures, these data are available consistently across large geographic areas, updated at relatively high frequencies, and can be aggregated to finer administrative units. 

The use of NTL as an indicator of economic activity is particularly well established. Henderson et al. \cite{Henderson2012}, for example, demonstrated that satellite-observed luminosity contains information that can supplement conventional income measures, particularly in contexts where statistical coverage is limited. Subsequent advances in satellite imagery, computational capacity and geospatial datasets have considerably expanded the range of observable characteristics, allowing researchers to combine measures of economic activity with information on built structures, connectivity, population concentration and environmental assets.

However, the increasing availability of such indicators raises a more fundamental question for composite measurement: what underlying concept of economic progress should these variables collectively represent? 

\subsection{Conceptualising Economic Progress} 
Economic progress is frequently equated with changes in output or income, whereas several contemporary approaches to welfare and wealth measurement distinguish current economic outcomes from the stock of resources and enabling conditions that make those outcomes possible. These include institutions and infrastructure that support households and firms, and the spatial organisation of economic activity. 

The OECD Well-being Framework, for example, distinguishes current well-being from the resources required to sustain well-being over time, organised around economic, human, natural and social capital \cite{OECD2024}. Similarly, the World Bank's Changing Wealth of Nations framework emphasises that GDP captures current production but does not fully account for the underlying productive base of an economy. This productive base includes produced capital such as roads and other infrastructure, human capital, natural capital and financial assets \cite{WorldBank2024}. UNEP's Inclusive Wealth approach likewise argues for complementing flow-based measures such as GDP with a productive stock-based assessment of produced, human and natural capital, since these assets determine an economy's capacity to create and maintain well-being over time \cite{UNEP2023}. This distinction provides the conceptual foundation for the Economic Progress Index developed in this study.

These frameworks provide an important starting point for interpreting the Economic Progress Index (EPI), but the EPI is narrower than a comprehensive measure of wealth or well-being. Its interpretation follows from the information that can actually be observed through the final set of variables entering the PCA-based index. The index combines information on road connectivity, the availability of schools, hospitals and banking institutions, night-time luminosity, and the scale and density of population. 

The study's geospatial architecture accordingly treats road networks as indicators of physical and commercial connectivity, hospitals and schools as service infrastructure, and banks as formal financial nodes. 

Taken together, these indicators are best understood as capturing the productive and enabling capacity of a local economy, the concentration of people and economic infrastructure within it, and the intensity with which that economic space is being utilised. This interpretation draws not only on the literature on capital and wealth, but also on theories of agglomeration and economic geography.

These mechanisms can be particularly important in metropolitan and peri-urban areas, where network effects may arise from the joint concentration of population, connectivity, service infrastructure and economic establishments. A bank located within an active commercial cluster may facilitate a larger volume of transactions; educational institutions located within a thick labour market may interact more closely with employers; and road networks connecting multiple centres of production and consumption can enlarge the effective market accessible to firms. Roads, schools, hospitals and banks principally describe relatively persistent elements of local capacity and infrastructure, whereas luminosity provides a flow-like signal of how intensively the local economic and settlement system is being used. Night-time lights per capita further help distinguish the scale of aggregate luminosity from its intensity relative to the resident population. The EPI does not directly estimate these agglomeration externalities. Instead, it captures the spatial conditions within which such network effects can emerge.

\subsection{Study Contributions and the Economic Progress Index (EPI) Framework}
The study makes three primary contributions. First, it builds a subdistrict-level socio-economic index entirely from geospatial data, without relying on official accounts or household surveys. Indicator weights come from Principal Component Analysis (PCA), so they reflect the shared variation across indicators and the index is constructed by a method that is transparent, replicable and extendable to new regions and periods. Second, it strengthens the interpretability of PCA-based measurement. Standard PCA can assign negative weights to indicators that should raise development, such as roads or schools, which makes the index hard to read. Benchmarking standard PCA against a non-negative variant, the study shows that this constraint preserves ranking and predictive accuracy almost entirely, yielding an index that is both statistically robust and directly interpretable. Third, it tests how closely EO-derived indicators track outcomes conventionally measured through household surveys and official accounts: wealth, consumption and district output. Using Maharashtra as a pilot, the framework offers a scalable, price-invariant basis for high-frequency monitoring of regional development, and supports more targeted, evidence-based planning and resource allocation.

The objectives of the study are to: (i) construct a statistically robust subdistrict-level Economic Progress Index using EO and geospatial datasets; (ii) identify spatio-temporal disparities in economic progress across Maharashtra; (iii) validate the index against established socio-economic benchmarks; and (iv) assess the potential of alternate data sources to complement traditional survey and administrative systems for development monitoring.

\section{Related Work}
\label{sec:related_work}

The deployment of EO and geospatial analytics to develop and monitor a subnational socio-economic variance-based index operates at the intersection of four rapidly evolving domains: satellite-based economic proxies, spatiotemporal signal decomposition, multi-indicator aggregation paradigms, and high-frequency macroeconomic nowcasting. Constructing policy-defensible development trackers requires resolving critical methodological bottlenecks across each stage: from sensor-level noise to multi-collinearity in feature aggregation.

\subsection{Nighttime Lights and Single-Proxy Limitations}
Being strongly correlated with aggregate anthropogenic activity, satellite-observed Nighttime Light (NTL) became an established direct proxy for subnational GDP growth, urbanization, and energy consumption in remote sensing econometrics \cite{Henderson2012, Agnihotri2021}. NTL intensity is often considered as an objective, high-frequency alternative to traditional census surveys, particularly in developing economies or data-scarce environments. Henderson et al. \cite{Henderson2012} deployed the Defense Meteorological Satellite Program Operational Line-Scan System (DMSP-OLS) to argue that NTL, when available, must be used alongside GDP to obtain a more accurate measure of economic growth, especially for developing countries with a large informal sector. In the Indian context of high-frequency macroeconomic tracking, NTL has been successfully deployed to capture short-term output contractions and informal sector shocks at the district level \cite{Beyer2018, ChodorowReich2020}. 

However, legacy NTL platforms suffer from severe technical constraints that compromise fine-scale economic estimation. Chief among these is dynamic range saturation (top-coding), where DMSP sensors max out at a Digital Number (DN) value of 63 over bright urban cores, masking growth variations within prosperous cities. Additionally, coarse spatial resolution causes severe radiance blooming and spatial overglow, the scattering of light into adjacent unlit pixels, unpopulated rural zones, or coastlines. This overglow artificially expands urban footprints, blurs true boundary distinctions, and distorts micro-level welfare and poverty estimates \cite{Mathen2024, Yusoff2026}. While newer platforms such as the Visible Infrared Imaging Radiometer Suite (VIIRS) eliminate top-coding and offer significantly sharper spatial resolution, purely NTL-driven models retain fundamental analytical limitations.

Beyond hardware-specific artifacts, NTL-only frameworks are constrained by systemic socio-economic blind spots. Long-term temporal studies frequently grapple with inter-satellite calibration inconsistencies across sensor transitions \cite{Fotopoulou2026}. More critically, nocturnal luminosity correlates poorly with agricultural productivity, land-use intensity, or non-illuminated rural poverty; economic dimensions that produce minimal light output yet represent substantial shares of subnational activity \cite{VoxDev2023, Fotopoulou2026}. As Asher et al. \cite{Asher2021} demonstrate across 600,000 Indian habitations in the SHRUG platform, while NTL correlates strongly with welfare in cross-sectional models, its time-series elasticity is dramatically lower, rendering single-proxy NTL tracking vulnerable to temporal drift. As a result, relying exclusively on NTL can lead to skewed regional economic assessments unless complemented by multi-spectral infrastructure indicators.

\subsection{Multispectral Fusion and Physical Asset Extraction}
To overcome these constraints, modern frameworks utilize integrated high-spatial-resolution multispectral imagery and computer vision to extract explicit structural proxies of wealth and land use, such as roof materials, road network density, building counts, building height proxies derived from shadow cast, and vehicle counts \cite{Engstrom2022}. Additionally, composite index architectures, such as the Vegetation--Water--Building--Thermal Nighttime Urban Index (VWBTNUI) developed by Yusoff et al. \cite{Yusoff2026} fuses NTL with spectral and thermal bands to eliminate light spillover and re-establish physical consistency in built-up area mappings. While composite indices like VWBTNUI demonstrate the value of multi-sensor fusion for urban mapping, their primary focus remains on spatial feature extraction rather than long-term spatiotemporal modeling. This paper addresses single-proxy NTL drift by integrating VIIRS luminosity within a multi-sensor fusion matrix, pairing nocturnal radiometry with daytime land-use, vegetation, and active infrastructure footprints to anchor economic signals in physical assets rather than light output alone.

In parallel, deep learning architectures applied to open-access multispectral imagery (e.g., Sentinel-2, Landsat) can explain up to 70\% of the variation in ground-measured village asset wealth across diverse sub-Saharan landscapes \cite{Yeh2020}. These spectral features extend to environmental, agricultural, and land-tenure dynamics. Studies demonstrate that principal component-derived spectral indices (such as the Standardized Spectral Reflectance Index, or SSRI) extracted from multispectral sensors significantly improve the spatial modeling of soil macronutrients and agricultural viability \cite{Dianati2025}, while fine-scale multispectral mapping enables precise quantification of green infrastructure deficits and agrarian structural shifts \cite{Engstrom2022, Catalan2026}. Within the Indian empirical context, computer vision applied to daytime imagery has successfully mapped rural poverty \cite{Das2020} and constructed village- and district-level Human Development Index (HDI) proxies across decadal census rounds \cite{Bansal2020, Gulgulia2023}.

\subsection{Sub-National Accounting Bottlenecks and Spatial Polarization}
Prior machine learning approaches (e.g., \cite{Das2020, Bansal2020}) rely heavily on ground-truth labels from infrequent national censuses (2001, 2011), subjecting models to temporal domain shifts over long time horizons while using opaque neural networks that lack policy-interpretable feature weights. Moreover, existing macroeconomic tracking remains predominantly anchored at the district level \cite{Beyer2018, ChodorowReich2020}, masking severe intra-district inequality at the subdistrict (taluka/tehsil) scale where administrative GDP labels do not exist. Recent institutional frameworks similarly emphasize the imperative to transcend administrative reporting aggregates; notably, Marini et al. \cite{Marini2026} in an IMF Working Paper demonstrate how integrating Earth Observations with spatial machine learning enables the generation of high-resolution gridded macroeconomic accounts, effectively downscaling national aggregates into localized economic footprints. Here, label dependency is bypassed entirely by constructing an unsupervised, fully interpretable feature matrix across 357 subdistricts of Maharashtra, combining passive remote sensing (LULC, LST, NDVI, NDBI) with active infrastructure footprints from OpenStreetMap without requiring survey training labels or black-box neural architectures.

This reliance on top-down district-level accounting is further challenged by severe sub-national mismeasurement and extreme spatial polarization. Conventional top-down District Domestic Product estimation systematically distorts ground realities, underestimating unincorporated manufacturing Gross Value Added (GVA) by 57.00\%--148.00\% due to its inability to capture informal sector throughput and localized shocks \cite{Seth2026}. Concurrently, formal economic output remains hyper-concentrated: in Maharashtra, just five urban growth poles; Mumbai, Thane, Pune, Nashik, and Nagpur, generate approximately 55.00\% of state GSDP, masking rural hinterland stagnation \cite{Kapoor2024}. Multi-decadal spatial econometric evaluations of Economic Census panels confirm that this intra-state industrial concentration is structurally entrenched, driven by high-technology manufacturing clusters that have remained geographically fixed for decades \cite{Tewari2025}. Together, these accounting bottlenecks and spatial asymmetries demonstrate why top-down administrative metrics fail, necessitating non-survey, high-resolution geospatial trackers at the subdistrict scale.

\subsection{Spatiotemporal Signal Decomposition}
In practice, unadjusted satellite time series confound long-term growth with seasonal phenology, cloud cover, and weather anomalies. Applying seasonality and trend decomposition using Locally Estimated Scatterplot Smoothing (LOESS) decouples time series into orthogonal trend, seasonal, and residual components \cite{Cleveland1990}. Isolating the trend before dimensionality reduction removes background noise, accounting for substantial (over 67\%) of error reduction in forecasting models \cite{Liu2025}. This study deploys a parameter-optimized Seasonal and Trend decomposition using LOESS (STL) ($w = 7$) across an 11.5-year panel (2015--2026) in Maharashtra, India, filtering monsoon cloud artifacts and agricultural harvest seasonality, ensuring the EPI isolates true structural capital accumulation. 

\subsection{Multi-Indicator Aggregation Architectures}
Principal Component Analysis (PCA) is widely used to construct area-level socioeconomic indices by isolating dominant variance axes without subjective indicator weighting \cite{Vyas2006, Krishnan2010, Jamalfaisal2025}. However, because unconstrained PCA is rotationally invariant, it frequently yields negative loadings for fundamentally positive assets, distorting index interpretability and penalizing genuine development gains. Rather than resorting to heuristic sign reversals or alternative weighting schemes that abandon variance maximization, this study adopts Non-negative PCA \cite{Montanari2015}. By restricting principal vectors to the positive orthant, Non-negative PCA guarantees strictly positive, policy-interpretable factor weights. Under high-dimensional spiked covariance regimes where the number of features grows proportionally with the sample size, Montanari and Richard \cite{Montanari2015} demonstrate that this constraint lowers the fundamental signal-to-noise detection threshold compared to classical PCA, facilitating reliable signal recovery in noisy, data-scarce settings. Moreover, while exact non-negative eigenvalue problems are non-convex and NP-hard, near-optimal solutions can be computed in near-linear time via Approximate Message Passing algorithms, providing a statistically sound, computationally scalable aggregation architecture for high-dimensional geospatial panels. 

\subsection{Unsupervised Spatial Nowcasting}
Spatial nowcasting uses machine learning and state-space models on EO data to estimate subnational GDP prior to official releases, bypassing publication lags and spatial aggregation loss \cite{Fotopoulou2026, Marini2026}. These enable spatially granular analyses that can help uncover pronounced spatial disparities and compounding vulnerabilities in data-scarce environments for precision policy targeting, which can be leveraged for developing scalable and actionable insights into poverty alleviation\cite{Mandon2026}. While recent institutional frameworks deploy machine learning to produce gridded macroeconomic outputs from satellite data \cite{Marini2026}, supervised nowcasting models still fail at hyper-local scales where official localized productivity training labels do not exist, which is typically the case in data-poor developing regions. We address this limitation by building an unsupervised, STL-adjusted subdistrict index and validating its structural legitimacy against official District GDP across four consecutive financial years, offering a reliable framework for monitoring hyper-local inequality where official accounts fall silent. Together, these new methods move beyond single-measure approaches to multi-sensor systems that can tell the difference between long-term growth and short-term disruptions.

\section{Methods}
\label{sec:methods}

\subsection{Study Design Rationale}

\subsubsection{Dual-Panel Architecture (2015--2026)}
To rigorously evaluate the proposed Economic Progress Index (EPI), this study adopts a dual methodological architecture comprising a 46-quarter longitudinal panel (2015 Q1--2026 Q2; $n = 16,422$ across 357 subdistricts) and a dedicated cross-sectional reference slice (2024). This dual design addresses two distinct methodological requirements.

The cross-sectional slice isolates a clean structural baseline comparison against ground-truth survey benchmarks provided by static, periodic datasets like National Family Health Survey (NFHS-5) Wealth Index and the NSSO Household Consumption Expenditure Survey (HCES) 2023--24, and the official real district GDDP without temporal mismatch decay. However, static cross-sections cannot evaluate whether an index tracks genuine structural development or merely reflects transient, seasonal spikes and atmospheric noise. An index constructed via a 46-quarter panel allows for non-parametric STL filtering, proving that it maintains low temporal rank volatility over time while remaining sensitive to local macro-economic development momentum. This longitudinal panel helps build operational reliability and facilitate nowcasting. 

The year 2024 was selected as the reference cross-sectional slice based on three criteria: firstly, it provides the closest temporal alignment with the most recent official ground-truth datasets available in India, specifically the 2023--24 HCES estimates and 2023--24 District Domestic Product accounts. Second, 2024 represents a post-pandemic relative structural equilibrium, avoiding the severe exogenous distortions of the 2020--2022 COVID-19 lockdown and recovery phases. Third, it offers peak spatial completeness for open-source vector infrastructure layers (OpenStreetMap) and VIIRS Nighttime Lights V2.1 collections, minimizing mapping completeness bias across remote agrarian subdistricts.

The period 2015--2026 was chosen to leverage the most out of the modern era of high-resolution multispectral EO data and serves three major advantages. First, the European Space Agency (ESA) launched Sentinel-2A in 2015 (with consistent data obtained from 2017), and the Suomi-NPP VIIRS nightlight monthly series stabilized its stray-light corrected processing pipelines starting around 2014--2015, providing consistent high-quality nocturnal luminosity data from 2015 onward. Second, the year 2015 marks the initiation of major national digital, financial, and physical infrastructure initiatives in India (such as Digital India launched in July 2015, widespread financial inclusion under PMJDY, and accelerated PMGSY rural road construction). Starting in 2015 Q1 captures the baseline prior to these major spatial infrastructure transformations, providing a policy and infrastructural baseline. Third, official district GDP (GDDP) and census-style survey accounts lag real-time economic activity by 2 to 3 years (currently available up to FY24/FY25). Extending the panel up to 2026 Q2 proves the practical spatial nowcasting utility of the index, demonstrating that the EPI can monitor live, present-day subnational economic trajectories where official accounts are yet to catch up.

\subsubsection{Quarterly Cadence and Phenological Filtering}
A quarterly cadence, structured strictly around standard calendar year quarters (Q1: Jan--Mar through Q4: Oct--Dec), resolves fundamental atmospheric, phenological, and econometric trade-offs. Annual metrics are too coarse to isolate seasonal noise, whereas monthly optical series in much of India are severely degraded by monsoon cloud cover (June--September). A three-month composite window guarantees sufficient cloud-free scene availability for high-fidelity spatial mosaics while cleanly isolating the phenological duality of India's Kharif and Rabi agricultural cycles ($p = 4$). Methodologically, quarterly aggregation acts as a physical low-pass filter: dampening transient surface noise (e.g., rain-fed greenness) to capture true capital accumulation (e.g., built-up expansion). Crucially for policy end-users, calendar quarters align directly with the quarterly release cadences of the Ministry of Statistics and Programme Implementation (MoSPI) and state economic bureaus, enabling government officials to conduct direct, real-time nowcasting without temporal interpolation errors.

\subsubsection{Pilot Study Context: Maharashtra}
The state of Maharashtra was chosen to pilot the index due to its unique characteristics and advantages. Located in western India, it is the country's second-most populous state and one of its largest contributors to national economic output. The state publishes district-level domestic product data annually (GDDP) and exhibits substantial spatial heterogeneity, encompassing highly urbanized metropolitan regions such as Mumbai, Pune, and Nagpur, alongside predominantly agrarian, tribal, forested, and drought-prone regions. This diversity in economic structure, data and infrastructure availability, environmental conditions, and settlement patterns makes Maharashtra an appropriate sandbox setting for evaluating the ability of geospatial indicators to capture subdistrict-level variation in economic progress.

\subsection{Methodological Workflow}
The EPI was developed as a composite measure of subdistrict-level economic progress using EO, geospatial, and demographic datasets. The methodological framework comprised five stages: (i) data acquisition and spatial harmonization, (ii) feature engineering and standardization, (iii) index construction through alternative weighting schemes, (iv) time-series decomposition (only for long-term analyses), and (v) external validation against established socio-economic benchmarks.

\subsubsection{Data Sources and Spatial Harmonization}
The analysis was conducted for all subdistricts in Maharashtra using the Socioeconomic High-resolution Rural-Urban Geographic Platform for India (SHRUG) v2.1 administrative boundaries. 

To measure productive capacity (structural productive base and assets) rather than transient transactional flows (such as short-term quarterly market turnover), indicators must reflect produced physical capital, service-enabling infrastructure and access nodes, and agglomeration scale. The multi-source satellite datasets used in this study consist of Earth Observation, Vector Networks, Gridded Population, and Climate and Land-based indicators. 

All datasets were aggregated to subdistrict boundaries obtained from SHRUG v2.1. Raster datasets were processed using zonal statistics to derive summary measures within each administrative unit, while vector datasets were converted into density- and accessibility-based indicators. Table~\ref{tab:datasets} summarizes the data sources used in the analysis.

\begin{table}[htbp]
\centering
\caption{Geospatial Datasets used for Index Construction}
\label{tab:datasets}
\small
\begin{tabularx}{\textwidth}{@{} >{\raggedright\arraybackslash}p{4.8cm} >{\raggedright\arraybackslash}X r @{}}
\toprule
\textbf{Indicator Variable} & \textbf{Source Dataset} & \textbf{Spatial Resolution} \\
\midrule
\texttt{VIIRS\_PER\_CAPITA\_HRSL}: Radiance per inhabitant & VIIRS Day/Night Band V2.1 & $\sim$500\,m \\
\addlinespace
\texttt{MEAN\_NDVI}: Vegetation;\newline \texttt{MEAN\_NDBI}: Built structure & Sentinel-2 Surface Reflectance (L2A) & 20\,m \\
\addlinespace
\texttt{LST\_MEAN}: Mean land surface temperature & MODIS LST (MOD11A2.061) & 1\,km \\
\addlinespace
\texttt{MEAN\_BARE}, \texttt{MEAN\_CROPS}: Land cover fractions & Dynamic World / Landsat Collection 2 & 10\,m \\
\addlinespace
\texttt{CHIRPS\_MEDIAN\_MM}: Median quarterly rainfall depth & CHIRPS Daily v2.0 & $\sim$5.5\,km \\
\addlinespace
\texttt{ROAD\_DENSITY} & OpenStreetMap (OSM) & Vector \\
\addlinespace
\texttt{POP\_DENSITY\_HRSL}: Population density & High Resolution Settlement Layer & $\sim$30\,m \\
\addlinespace
\texttt{HOSPITALS}, \texttt{BANKS}, \texttt{SCHOOLS} per 10k people & OpenStreetMap (OSM) & Vector \\
\bottomrule
\end{tabularx}
\end{table}

\subsubsection{Feature Engineering and Normalization}
A series of preprocessing and feature engineering steps were undertaken to ensure comparability across datasets with differing spatial resolutions, temporal frequencies, and measurement units.

For the cross-sectional analyses, raster datasets, annual composites were generated to reduce seasonal fluctuations and minimize the influence of transient observations. Zonal statistics, including mean, median, and aggregate values, were computed within each subdistrict polygon. 

For the longitudinal analyses across 46 quarters, features were structured across four normalization tiers: (i) Population-based: HRSL population density was scaled by land area to estimate total subdistrict population, enabling the derivation of per-capita nighttime luminosity and OpenStreetMap (OSM) point-amenity intensity per 10,000 people; (ii) Area-based: OSM transport layers were accounted for to compute road density in kilometres; and (iii) Zonal aggregations: Means of continuous raster products were extracted as spatial zonal averages within SHRUG v2.1 subdistrict boundaries. (iv) Additionally, district, revenue division, and state-level averages of the indices have been weighted by HRSL population data to ensure representative regional aggregation.

To account for skewed distributions, all indicators with an absolute skewness greater than 1 were transformed using $\log(x+1)$ prior to standardization. To facilitate aggregation across heterogeneous indicators, all variables were standardized using $z$-score normalization:
\begin{equation}
z_i = \frac{x_i - \bar{x}}{s}
\end{equation}
where $x_i$ denotes the observed value, $\bar{x}$ the sample mean, and $s$ the sample standard deviation. Indicators negatively associated with economic progress, such as elevated land surface temperature or environmental stress measures, were directionally transformed so that higher values consistently reflected more favorable development outcomes.

\subsubsection{Time-Series Decomposition and Trend Extraction}
For the 46-quarter longitudinal panel, quarterly values obtained upon imputation were subjected to STL decomposition to decouple long-term structural development from recurring seasonal phenology and acute exogenous shocks \cite{Cleveland1990}. The trend component thereby extracted was used as the input separately for both Standard and Non-negative PCA to extract their respective first principal components.

\begin{figure}[htbp]
\centering
\includegraphics[width=1\textwidth]{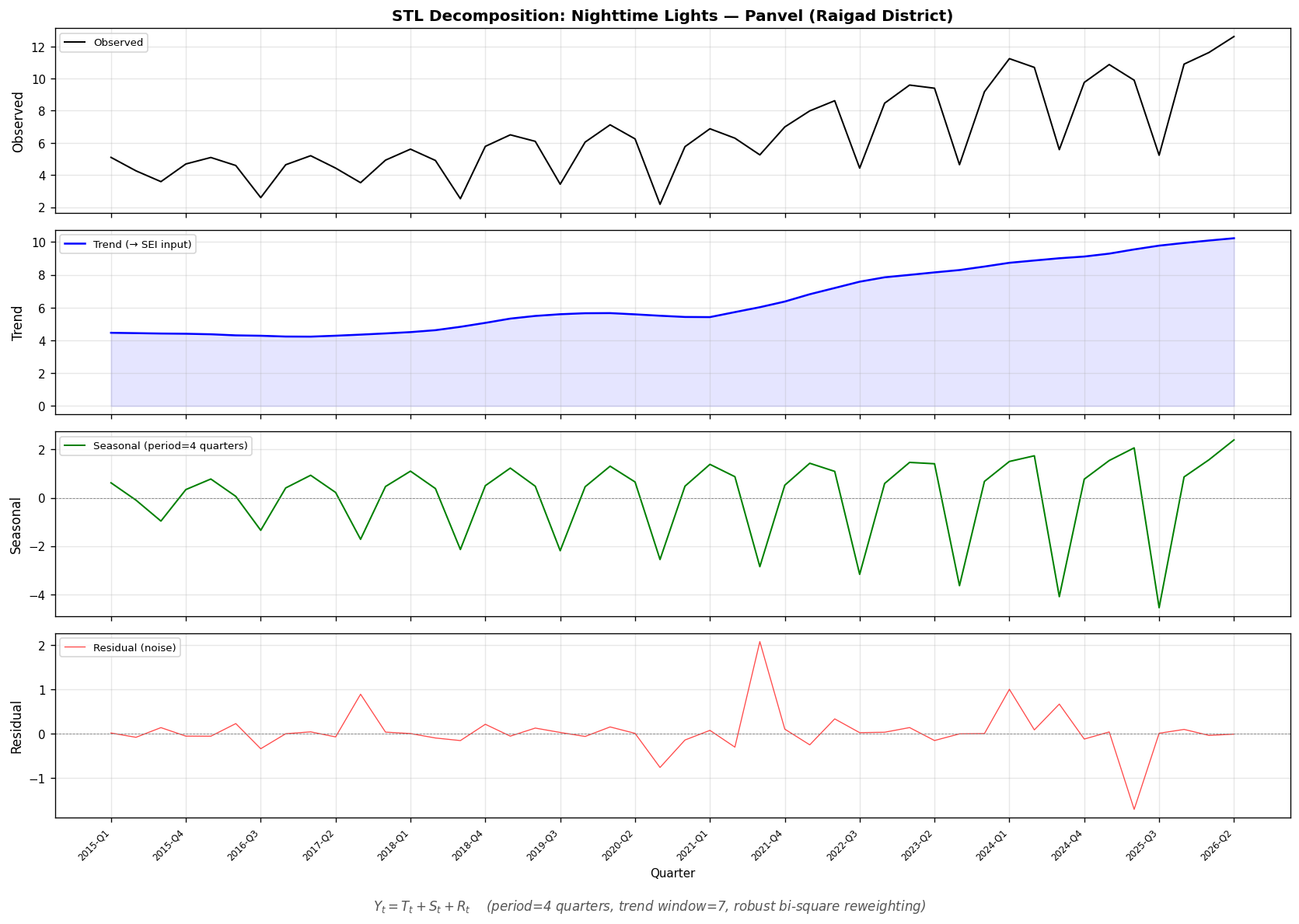}
\caption{STL Decomposition of Panvel Subdistrict from 2015 Q1 to 2026 Q2.}
\label{fig:stl}
\end{figure}

STL decomposes each series into three orthogonal components: a low-frequency trend, a periodic seasonal pattern, and an irregular residual (Figure~\ref{fig:stl}). To prevent severe, non-periodic dislocations and exogenous shocks (such as the 2020 COVID-19 mobility restrictions and extreme monsoonal anomalies) from leaking into the residual component, we specified a flexible LOESS trend window parameter of seven quarters ($w_t = 7$). While standard automated defaults ($w_t \ge 13$) produce overly rigid trend components that displace transient structural drops into residual noise \cite{Hyndman2021}, a 7-quarter span ($\approx 1.75$ years) strikes an optimal balance: it allows the trend line to bend during multi-quarter macroeconomic shocks, while robust bi-square weighting confines isolated single-quarter spike outliers in the residual series, preventing them from pulling historical trend estimates out of shape.

While X-13ARIMA-SEATS remains the institutional standard for seasonal adjustment in official national accounts, STL offers critical methodological advantages for processing this multi-sectoral geospatial panel dataset. First, its non-parametric flexibility allows it to handle highly non-linear environmental and climate signals (such as NDVI and LST) without requiring rigid parametric assumptions or the complex, manual specification of individual ARIMA models for each of the 357 subdistricts. Second, STL provides exceptional computational scalability, avoiding the high processing times and model convergence failures inherent in automated ARIMA estimation over large spatial grids. Furthermore, unlike the relatively rigid filters utilized in X-13ARIMA, STL features dynamic seasonal adaptation, allowing seasonal amplitudes to evolve fluidly over the 11.5-year horizon to capture shifting monsoon patterns and climate variations. 

Despite these advantages, caveats specific to the STL framework need to be addressed. STL lacks automatic calendar adjustments for moving holidays or festivities (like Diwali or Eid), and the trend window selection is empirical, yet pragmatic. Furthermore, because LOESS relies on local data neighborhoods, trend estimations at the series boundaries (2015 Q1 and 2026 Q2) are less stable and prone to revisions when new data is added. However, these boundary variations can be programmatically stabilized in future iterations once the EPI transitions to a long-term, rolling dynamic calculation framework. 

\subsubsection{Composite Index Aggregation Architectures}
The Economic Progress Index was constructed using two weighting approaches involving Principal Component Analysis (PCA): Standard PCA and Non-Negative PCA. Comparing these approaches enabled an assessment of how different assumptions regarding indicator correlation influence composite index performance. PCA identifies latent dimensions of variation across subdistricts and transforms correlated variables into orthogonal principal components that successively maximize explained variance. The first principal component (PC1), capturing the maximum proportion of total spatial variance, was extracted to construct the weight vector.

Prior to weight derivation, all features were directionally aligned so that higher values consistently indicate positive developmental progress, ensuring policy interpretability and structural consistency. 

Let $\mathbf{Z} \in \mathbb{R}^{N \times p}$ denote the standardized feature matrix, with entry $z_{j,i}$ for subdistrict $j \in \{1,\dots,N\}$ and indicator $i \in \{1,\dots,p\}$, and $\mathbf{S} = \frac{1}{N-1}\mathbf{Z}^T \mathbf{Z}$ denote the sample covariance matrix.

The eigenvalues ($\lambda$) and $L_2$-normalized eigenvectors ($\mathbf{v}$) of $\mathbf{S}$ are calculated by solving the characteristic equation:
\begin{equation}
\mathbf{S}\mathbf{v} = \lambda \mathbf{v}
\end{equation}
The first Principal Component (PC1) corresponds to the $L_2$-normalized eigenvector $\mathbf{v}_1 = (v_{1,1}, \dots, v_{p,1})^T$ associated with the largest eigenvalue, $\lambda_1$. 

\paragraph{Standard PCA}
The unit normalized loadings/weights matrix ($\mathbf{W} = [w_i] \in \mathbb{R}^p$) associated with PC1 is given by:
\begin{equation}
\label{eq:std_pca_weights}
w_i = v_{i,1}
\end{equation}
The final PCA-based Economic Progress Index score for subdistrict $j$ was calculated as:
\begin{equation}
\mathrm{EPI}_j^{\mathrm{PCA}} = \sum_{i=1}^p w_i z_{j,i}
\end{equation}

\paragraph{Principal Component Analysis (PCA) and Non-Negative Optimization}
Standard PCA allows unconstrained eigenvectors ($\mathbf{v} \in \mathbb{R}^p$), which can assign negative loadings to variables that inversely correlate with the dominant development axis. To enforce structural non-negativity and eliminate artificial penalties on non-urban features, we solve the constrained non-negative PCA optimization problem \cite{Montanari2015}:
\begin{equation}
\mathbf{v}^* = \arg\max_{\mathbf{v}} \left( \mathbf{v}^T \mathbf{S} \mathbf{v} \right) \quad \mathrm{s.t.} \quad \mathbf{v} \ge 0, \; \|\mathbf{v}\|_2 = 1
\end{equation}
Since this non-convex quadratic program operates on the intersection of the non-negative orthant and the unit sphere, optimal loadings are computed via Projected Power Iteration:
\begin{enumerate}
\item \textbf{Initialization:} Set $\mathbf{v}^{(0)} = \frac{1}{\sqrt{p}} \mathbf{1}_p$ (uniform non-negative vector).
\item \textbf{Power Step:} Compute unconstrained candidate vector $\tilde{\mathbf{v}} = \mathbf{S}\mathbf{v}^{(t)}$.
\item \textbf{Projection:} Project onto the non-negative orthant $\mathbf{v}_+ = \max(\tilde{\mathbf{v}}, 0)$ element-wise.
\item \textbf{Normalization:} Update $\mathbf{v}^{(t+1)} = \frac{\mathbf{v}_+}{\|\mathbf{v}_+\|_2}$.
\item \textbf{Convergence:} Repeat until $\|\mathbf{v}^{(t+1)} - \mathbf{v}^{(t)}\|_2 < 10^{-10}$ (converges in $\sim$20 iterations for the dataset used in this study).
\end{enumerate}

\paragraph{Karush-Kuhn-Tucker (KKT) Zeroing Mechanics}
Under the Lagrangian \\$\mathcal{L}(\mathbf{v}, \lambda, \boldsymbol{\mu}) = \mathbf{v}^T \mathbf{S} \mathbf{v} - \lambda (\|\mathbf{v}\|_2^2 - 1) + \boldsymbol{\mu}^T \mathbf{v}$ with dual multipliers $\mu_i \ge 0$, \\complementary slackness dictates that for each indicator $i$:
\begin{equation}
(\mathbf{S}\mathbf{v})_i = 
\begin{cases}
\lambda v_i, & \text{if } v_i > 0, \\
\le 0, & \text{if } v_i = 0.
\end{cases}
\end{equation}
An indicator is strictly zeroed ($v_i = 0$) when its net covariance with the weighted development combination is negative, rather than receiving an arbitrary negative weight as in Standard PCA.

The final normalized weight for indicator $i$ is defined as $w_i = \frac{v_i}{\sum_{k=1}^p v_k}$, satisfying $\sum_{i=1}^p w_i = 1$. 

The EPI score for subdistrict $j$ is calculated as:
\begin{equation}
\mathrm{EPI}_j^{\mathrm{NN\text{-}PCA}} = \sum_{i=1}^p w_i z_{j,i}
\end{equation}

Because Standard PCA loadings are $L_2$-normalized (Eq.~\ref{eq:std_pca_weights}) whereas NN-PCA weights are $L_1$-normalized to sum to unity, raw composite scores from the two architectures lie on different arbitrary scales. To place both indices on a common dimensionless $[0,1]$ interval for cross-method comparison and spatial mapping, raw scores are linearly rescaled via global min--max normalization across all $16,422$ subdistrict-quarter observations:
\begin{equation}
\widetilde{\mathrm{EPI}}_{j,t} = \frac{\mathrm{EPI}_{j,t} - \min(\mathrm{EPI})}{\max(\mathrm{EPI}) - \min(\mathrm{EPI})},
\label{eq:minmax}
\end{equation}

A single pair of bounds applies to every quarter, so scores are comparable over time. The value 1 marks the highest subdistrict-quarter in the panel, and 0 the lowest. The same procedure is applied separately to the Standard PCA and NN-PCA composites.

District, division, and state values are population-weighted means of subdistrict scores, 
\begin{equation}
    \mathrm{EPI}_{k,t} = \sum_{j \in k} \pi_j \mathrm{EPI}_{j,t}
\end{equation}
 Where $\pi_j$ is subdistrict~$j$'s share of unit~$k$'s HRSL population. Financial-year values are the mean of the four quarterly trend values from April to March.

\subsubsection{Validation and Uncertainty Assessment}
The validity of the cross-sectional Economic Progress Index (EPI) was assessed against independent socio-economic benchmarks that were not used in its construction. External validation drew on three datasets capturing complementary dimensions of economic well-being: (i) the Wealth Index and the prevalence of clean cooking fuel use from the fifth National Family Health Survey (NFHS-5, 2019–21), which reflect household asset ownership and living standards; (ii) district-level Monthly Per Capita Consumption Expenditure (MPCE) estimates from the Household Consumption Expenditure Survey (HCES, 2023–24); and (iii) district-level real Gross Domestic Product (GDDP) estimates for 2023–24. These datasets were reserved exclusively for post hoc evaluation and played no role in index construction.

Validation was undertaken through three complementary approaches. First, rank-order comparisons were performed to assess the extent to which the Standard and Non-negative PCA-derived indices reproduced established spatial development hierarchies observed in the benchmark datasets using Spearman and Kendall coefficients. District rankings generated by the EPI were compared with corresponding rankings derived from NFHS wealth scores, HCES MPCE estimates, and GDP estimates.

Second, correlation analyses were conducted to quantify associations between the constructed indices and survey-based measures of socio-economic welfare. Pearson correlation coefficients were computed to evaluate linear relationships between index scores and benchmark indicators, while Spearman rank correlations were used to assess consistency in relative rankings. In addition to aggregate wealth and consumption measures, correlations were examined against selected NFHS-5 indicators, including access to sanitation, electricity, and clean cooking fuel, to evaluate the extent to which the EPI captured underlying dimensions of household welfare.

Third, predictive validity was assessed using Ordinary Least Squares (OLS) regression models. Bivariate regressions were estimated with NFHS-5 measures as dependent variables and the constructed EPI scores as predictors. Model performance was evaluated using the coefficient of determination ($R^2$) and Root Mean Squared Error (RMSE).

The internal reliability of the index was assessed through a spatial bootstrap confidence scoring framework. This approach incorporated spatial priors to account for geographic variation and data volatility across subdistricts, enabling differentiation between stable development signals and statistical noise arising from localized variability. Posterior confidence estimates were generated for each subdistrict and used to compare the stability of alternative weighting methodologies. Indicator weights were re-estimated in each resample. We report 95\% confidence intervals for the weights and a rank-confidence score for each subdistrict, defined as the share of resamples in which its rank moves by no more than ten positions from its full-sample rank. Together, these validation and uncertainty assessment procedures provided a comprehensive evaluation of both the external validity and internal reliability of the proposed index. 

Once the predictive accuracy of the spatial proxy layer is established, the longitudinal framework extends these validated indicators into a population-normalized, STL-filtered panel (2015 Q1--2026 Q2) to enable high-frequency temporal tracking without seasonal or demographic distortion. Longitudinal validation of the annual district-aggregated EPI was conducted against official Real District Gross Domestic Product (Real GDDP) accounts across four consecutive financial years (FY22--FY25; $n = 35 \times 4 = 140$ district-year observations). This time window was selected based on two empirical considerations. First, it is the period for which district estimates are available on a consistent methodological basis and excludes FY21, when lockdown-related disruptions severely distorted both economic activity and its measurement. Second, because district accounts are released with a substantial lag, the most recent years in this series are provisional/advance estimates subject to revision; validation against them therefore provides a conservative test of agreement with the latest official figures. The remainder of the panel (2025 Q2 --2026 Q2), for which official estimates are not yet available, illustrates the utility of the index to nowcast the current economic conditions.

\subsubsection{Convergence and Inequality Measures}
\label{sec:convergence_methods}

Regional inequality was assessed at two scales. At the revenue-division level, subdistrict EPI scores were aggregated to each of Maharashtra's six divisions using HRSL population weights. Relative ($\sigma$-) convergence was measured with the coefficient of variation across division means, $\mathrm{CV}_t = s_t / \bar{y}_t$, where $s_t$ is the sample standard deviation across divisions in quarter $t$. The trend in $\mathrm{CV}_t$ was estimated by OLS on a linear quarter index across the $46$-quarter panel. Absolute dispersion was measured with the standard deviation across divisions and the gap between the highest- and lowest-scoring divisions.

Conditional on initial levels, $\beta$-convergence was estimated as
\begin{equation}
g_i = \alpha + \beta \ln y_{i,0} + \varepsilon_i,
\label{eq:beta_convergence}
\end{equation}
where $g_i$ is the compound annual growth rate of unit $i$ between the baseline window (2015~Q1--Q4) and the latest window (2025~Q3--2026~Q2), and $y_{i,0}$ is its baseline EPI. The implied speed of convergence is $\lambda = -\ln(1 + \beta T)/T$, with $T \approx 10.5$~years between window midpoints, and the half-life is $\ln 2 / \lambda$. Both endpoints are four-quarter averages, which spans the seasonal cycle and reduces sensitivity to STL boundary estimates. Within-district inequality was measured as the CV of subdistrict scores within each district over the same windows.

\section{Results}
\label{sec:results}

\subsection{Economic Progress Index Results}

\subsubsection{Feature Weight Architecture: Standard and Non-Negative PCA}
The PCA-derived Economic Progress Index (EPI) captures the dominant spatial pattern of economic development across Maharashtra by assigning greater weight to correlated indicators associated with urbanization, infrastructure concentration, and economic activity. Consequently, the index functions primarily as a measure of structural development intensity.

For longitudinal panel data, applying non-negative PCA (NN-PCA) to the standardized matrix across 12 candidate indicators yielded a dominant first principal component explaining 45.4\% of total variance (compared to 46.1\% under standard PCA). Environmental, climate, and land-cover series; specifically rainfall, land surface temperature, vegetation index, cropland, and built-up index, exhibited near-zero or orthogonal loadings relative to physical development and were assigned zero weight under NN-PCA. The remaining 7 active variables in the NN-PCA-based EPI absorb the total index weight, concentrating primarily on physical transport networks and point-based public infrastructure.

\begin{figure}[htbp]
\centering
\includegraphics[width=1\textwidth]{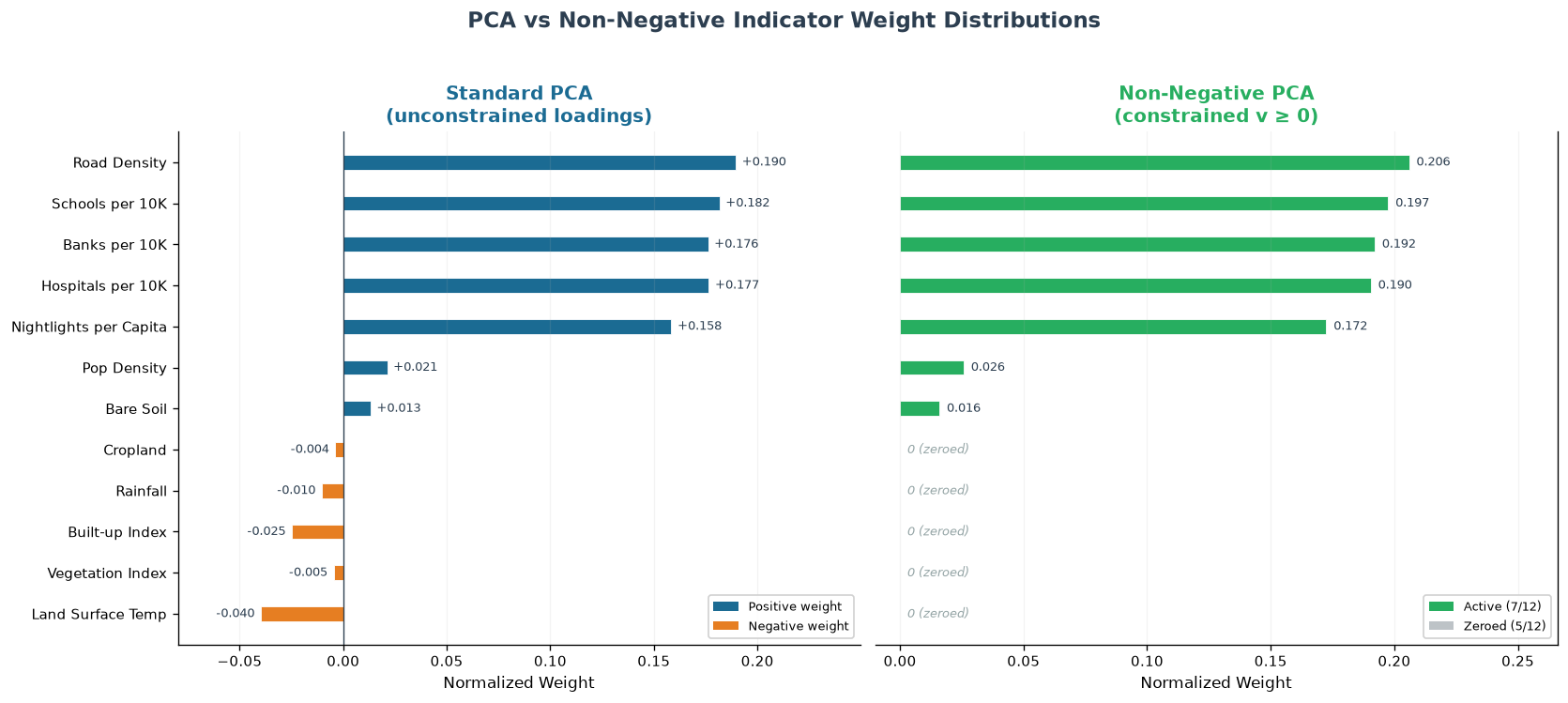}
\caption{PCA vs. Non-Negative PCA Indicator Weight Distributions. Derived weights (normalized) for indicators in Standard PCA (left, 12 variables, PC1 explaining 46.1\% of variance) and Non-negative PCA (right, 7 non-zero dimensions, PC1 explaining 45.4\% of variance).}
\label{fig:weights}
\end{figure}

\subsubsection{Rank Concordance and Spatial Displacement Dynamics}
\label{sec:rank_concordance}

Comparing Standard PCA and Non-Negative PCA (NN-PCA) clarifies how non-urban and environmental variables influence composite index levels without altering the overarching spatial hierarchy. Across the full 46-quarter panel ($n = 16,422$), the two weighting schemes yield virtually identical ordinal rankings (Spearman $\rho = 0.9981$, Kendall $\tau = 0.964$; rising to $\rho = 0.999$ in the 2026-Q2 cross-section), reflecting the dominance of physical connectivity and civic infrastructure loadings ($+0.158$ to $+0.190$; Figure~2). Because raw z-score composites are zero-centered and use different weight normalizations ($L_2$ vs.\ $L_1$), score dispersion is evaluated on the globally min--max rescaled $[0,1]$ indices (Eq.~\ref{eq:minmax}). Standard PCA exhibits a higher Coefficient of Variation ($CV = 0.709$) than NN-PCA ($CV = 0.581$). Rather than indicating a reduction in scale-invariant regional inequality, this lower CV reflects an upward shift of the lower rural tail relative to the panel minimum once small negative loadings; ranging from $-0.004$ for Cropland to $-0.040$ for Land Surface Temperature (Figure~2), are constrained to zero via KKT complementary slackness.

Although major metropolitan centers anchor the upper bound under both specifications, zeroing out the five negatively weighted variables (Rainfall, Land Surface Temperature, Built-up Index, Vegetation Index, and Cropland) produces modest, directionally consistent upward score adjustments in rural and agrarian regions---most visibly across interior divisions such as Amravati and Aurangabad. Because the negative loadings in Standard PCA are small in magnitude, their removal does not overhaul statewide rankings: across all $357$ subdistricts in 2026-Q2, the mean absolute rank shift between Standard PCA and NN-PCA is $2.80$ positions (median $2.0$, maximum $26$ positions; $69$ subdistricts shifting by $\ge 5$ positions), corresponding to a mean score increase of $+0.005$ normalized units in Amravati ($+0.006$) and Aurangabad ($+0.004$) during 2025-Q3--2026-Q2 (and $+0.018$ units during 2015-Q1 -- 2015-Q4). Thus, NN-PCA eliminates a subtle downward directional shift on ecological and agricultural land cover (as well as spectral noise from the Built-up Index) without distorting relative spatial rankings.

\begin{figure}[htbp]
\centering
\includegraphics[width=1\textwidth]{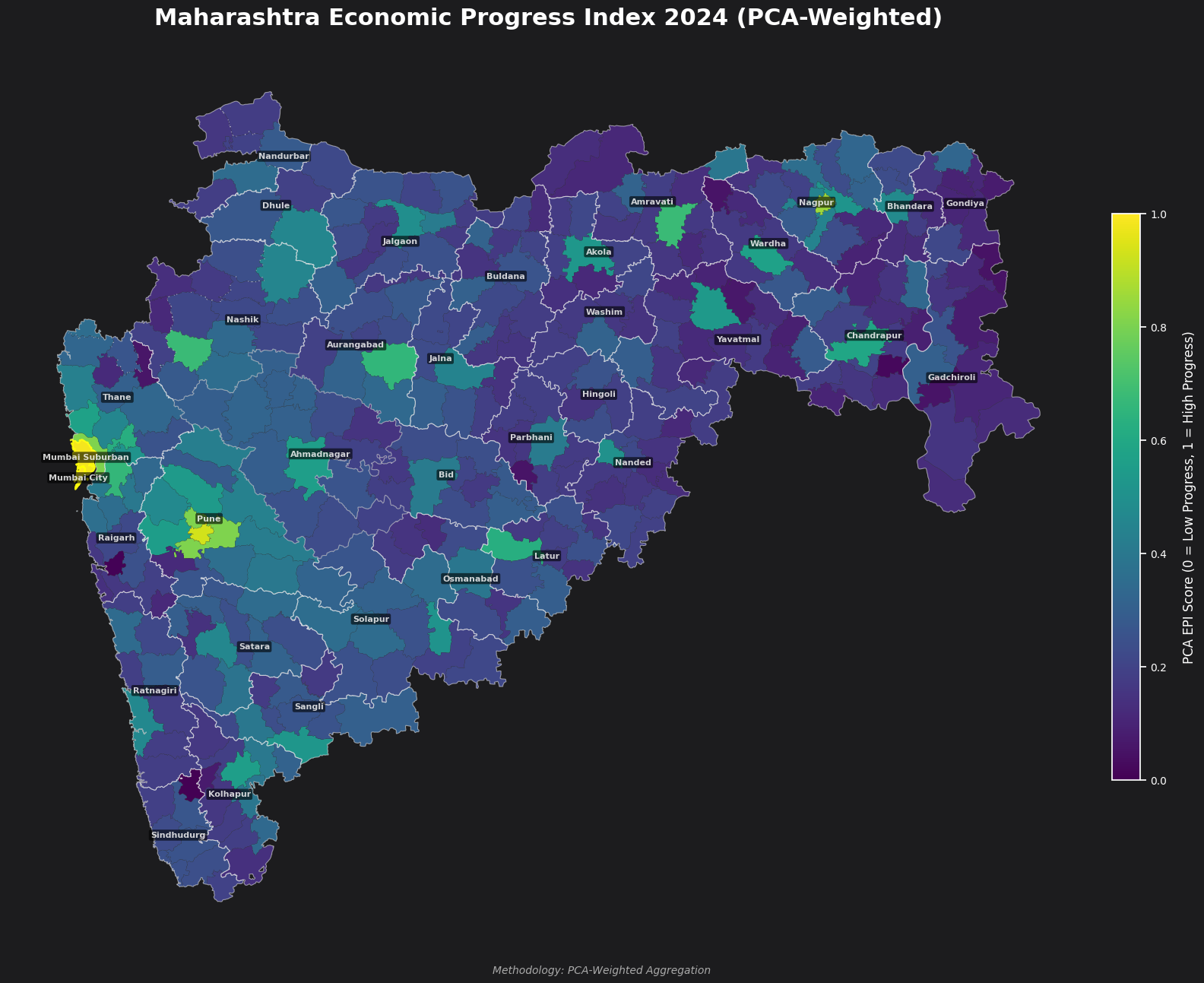}
\caption{Standard PCA-Weighted Economic Progress Index Distribution Across Maharashtra Subdistricts (2024).}
\label{fig:spatial_distribution}
\end{figure}

\subsubsection{Spatial Distribution of Infrastructure Agglomeration (PCA)}
\label{sec:spatial_distribution}

In the 2024 cross-sectional analysis, top metropolitan units; led by Mumbai Suburban, Pune City, Ulhasnagar, Nagpur, and Mumbai City, consistently define the upper bound of the min--max rescaled distribution ($\widetilde{\mathrm{EPI}} \to 1$; Eq.~\ref{eq:minmax}) across both Standard and Non-Negative PCA frameworks. 

These high-performing nodes are characterized by intense nighttime luminosity, dense transport connectivity, and concentrated commercial and industrial activity. Under the non-negative formulation ($\widetilde{\mathrm{EPI}}^{\mathrm{NN-PCA}}$), metropolitan centers establish the upper performance anchor without inducing negative-weight downward shifts in the lower tail of non-metropolitan administrative units.

Spatially, the index reveals a pronounced concentration of developmental activity within the primary Mumbai--Thane--Pune industrial corridor and regional growth poles such as Nagpur (Figure~\ref{fig:spatial_distribution}). Moving outward from these urban hubs, scores decline rapidly, exhibiting a steep developmental gradient across Maharashtra. Under Standard PCA, agricultural and forested subdistricts across Marathwada and Vidarbha (such as Sonpeth, Kalamb, Dhanora, and Vaijapur) experience modest downward score shifts driven by small negative loadings on land surface temperature, cropland, vegetation, and spectral-noise indices. Under NN-PCA, zeroing out these five indicators removes this downward tail shift while preserving underlying disparities in physical infrastructure density and nocturnal luminosity, faithfully reflecting lower urban economic intensity without treating rural land cover as an economic detriment.

The non-negative PCA framework also underscores substantial intra-district disparities. Major urban headquarters, such as Nagpur City, rank near the upper bound of the state distribution, whereas adjacent rural subdistricts within the same district exhibit markedly lower scores. This spatial divergence indicates that physical infrastructure and economic intensity remain concentrated within district headquarters and metropolitan corridors rather than diffusing evenly across district boundaries.

\subsubsection{Eliminating the Agrarian Penalty via Non-Negative Constraints}
Standard PCA assigns negative weights to five indicators---Land Surface Temperature, Cropland, Vegetation Index, Built-up Index, and Rainfall. This introduces a structural distortion: agrarian and forested subdistricts (such as those in Gadchiroli, Amravati, and Aurangabad) are actively penalized for high ecological or agricultural presence, artificially widening the urban-rural score gaps.

Non-Negative PCA (NN-PCA) resolves this through KKT complementary slackness conditions, which strictly zero out these 5 orthogonal indicators rather than subtracting them. By retaining 7 core infrastructure and economic indicators, NN-PCA eliminates artificial penalties on non-urban features while capturing a near-similar proportion of variance. 

Treating rural assets neutrally rather than negatively compresses overall inequality across districts and corrects the relative standing of rural subdistricts without suppressing metropolitan hubs. This non-negative formulation is empirically validated by better performance against independent benchmarks; improving explanatory power for both the NFHS Wealth Index ($R^2 = 0.743$) and NSSO Log MPCE ($R^2 = 0.516$), confirming that zeroing negative loadings removes structural bias rather than genuine economic signal.

\subsection{Validation Results}
\subsubsection{Internal Stability and Sampling Precision}
\label{sec:internal_stability}

To evaluate parameter precision and ordinal rank stability, we executed a $1,000$-iteration bootstrap resampling quarters with replacement within each of the $357$ subdistricts (Table~\ref{tab:bootstrap_stability}). Across resamples, Non-Negative PCA (NN-PCA) consistently achieves higher rank persistence and narrower rank uncertainty bounds than unconstrained Standard PCA while retaining $45.4\%$ of total variance (an efficiency ratio of $0.987$ relative to Standard PCA).

\begin{table}[htbp]
\centering
\caption{Bootstrap rank stability and confidence metrics across $1,000$ within-subdistrict quarterly resamples ($N = 357$ subdistricts; tolerance window $\pm 10$ rank positions).}
\label{tab:bootstrap_stability}
\begin{tabular}{lcc}
\toprule
\textbf{Metric} & \textbf{Standard PCA} & \textbf{NN-PCA} \\
\midrule
Mean rank confidence, $P(|\Delta \mathrm{rank}| \le 10)$   & 0.654  & 0.669  \\
Median rank confidence, $P(|\Delta \mathrm{rank}| \le 10)$ & 0.613  & 0.636  \\
Subdistricts with confidence $> 0.80$ (\%)                 & 25.8\% & 27.5\% \\
Subdistricts with confidence $> 0.50$ (\%)                 & 72.8\% & 77.0\% \\
Mean $95\%$ rank CI width (positions)                      & 45.9   & 44.1   \\
Median $95\%$ rank CI width (positions)                    & 47.0   & 45.0   \\
\bottomrule
\end{tabular}
\end{table}

\paragraph{Parameter Precision}
In Standard PCA, unconstrained estimation allows negative loadings on climate and land-cover indicators across resamples, introducing sign reversals and wide confidence bounds (Figure~4). Under NN-PCA, projected power iteration enforces non-negativity ($v \ge 0$), anchoring the core physical infrastructure and nocturnal luminosity weights within tighter bootstrap intervals and preventing negative-weight distortions when temporal quarters are resampled \cite{Montanari2015,Mathen2024}.

\paragraph{Rank Confidence and Boundary Anchoring}
Under NN-PCA, the mean subdistrict rank-confidence score ($P(|\Delta \mathrm{rank}| \le 10)$) reaches $0.669$ (median $0.636$), compared with $0.654$ (median $0.613$) under Standard PCA (Table~\ref{tab:bootstrap_stability}). Over three-quarters ($77.0\%$) of all subdistricts achieve a rank-confidence score above $0.50$ under NN-PCA (versus $72.8\%$ under Standard PCA), and $27.5\%$ exceed $0.80$ (versus $25.8\%$). Similarly, NN-PCA narrows the mean $95\%$ rank confidence interval width from $45.9$ to $44.1$ positions (and the median width from $47.0$ to $45.0$ positions). As illustrated in Figure~5, rank confidence follows a distinct U-shaped profile across the state hierarchy: metropolitan leaders at the upper bound and remote rural subdistricts at the lower bound exhibit near-unity rank stability, whereas mid-tier subdistricts---where baseline infrastructure endowments are closely clustered---account for most within-band rank movement \cite{Montanari2015,OECD2024}.

\begin{figure}[htbp]
\centering
\includegraphics[width=1\textwidth]{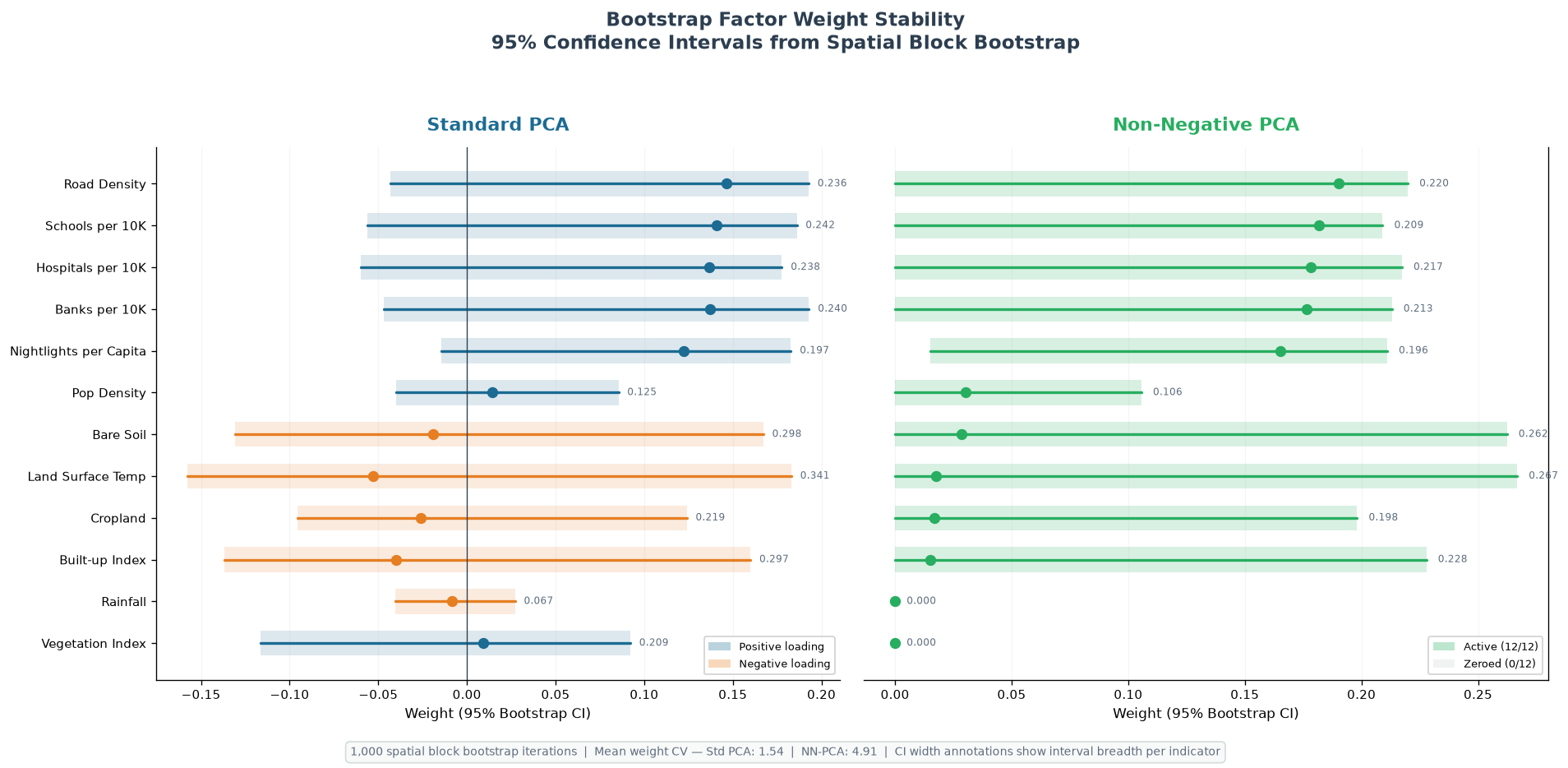}
\caption{Bootstrap Factor Weight Stability. Parameter-level weight stability under a 95\% bootstrap confidence interval.}
\label{fig:bootstrap}
\end{figure}

\begin{figure}[htbp]
\centering
\includegraphics[width=1\textwidth]{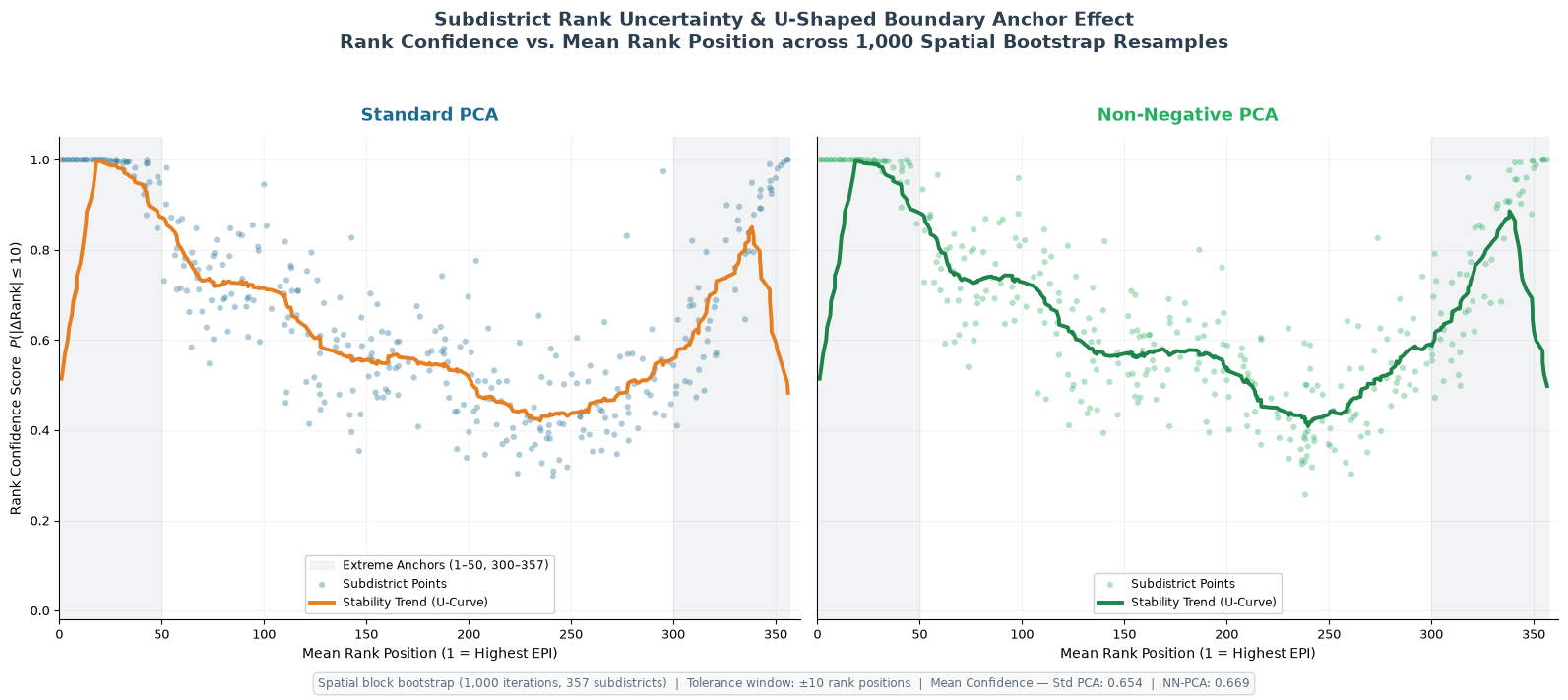}
\caption{Subdistrict Rank Uncertainty Analysis. Rank position vs. confidence score, demonstrating the U-shaped boundary effect.}
\label{fig:rank_uncertainty}
\end{figure}

\subsubsection{Longitudinal Panel Dynamics and Variance Decomposition}
Tracking $N = 357$ subdistricts across the 46-quarter panel demonstrates high temporal persistence and minimal short-term rank jitter. While Standard PCA's negative loadings on seasonal variables (Rainfall, Land Surface Temperature, Vegetation Index, Built-up Index, and Cropland) introduce minor phenological rank volatility, Non-Negative PCA (NN-PCA) achieves comparable temporal stability.

Decomposing panel variance using a Two-Way Fixed Effects specification ($\mathrm{EPI}_{i,t} = \alpha_i + \gamma_t + \epsilon_{i,t}$) shows that spatial structural baselines ($\alpha_i$) account for over 91\% of total variance in the NN-PCA model. Statewide quarterly trends ($\gamma_t$) account for macro-economic growth trajectories, while unexplained residual noise ($\epsilon_{i,t}$) drops below 5\% of total variance. This yields a high Signal-to-Noise Ratio ($\mathrm{SNR} > 12.0$) for NN-PCA, outperforming Standard PCA. Operating on STL-decomposed trend components ($\mathrm{period} = 4, \mathrm{seasonal} = 7$) and constraining counter-intuitive negative loadings to zero, NN-PCA naturally insulates the index from seasonal atmospheric and agricultural fluctuations without distorting long-term spatial baseline relationships.

\subsubsection{External Survey and Macroeconomic Validation}
To evaluate external validity, both Non-Negative PCA (NN-PCA) and Standard PCA specifications were benchmarked against independent survey measures of socio-economic well-being; the NFHS-5 Wealth Index and the NSSO Household Consumption Expenditure Survey (HCES 2023--24 Log MPCE), at the district level using population-weighted averages.

Across both benchmarks, NN-PCA is consistently comparable to Standard PCA, demonstrating that enforcing non-negativity ($\mathbf{v} \ge 0$) and zeroing out 5 non-urban indicators removes noise without sacrificing predictive capacity:
\begin{itemize}
\item \textbf{NFHS-5 Wealth Index:} 

Table~\ref{tab:nfhs_ols} reports bivariate Ordinary Least Squares (OLS) regressions of district-level National Family Health Survey (NFHS-5) mean wealth index scores on the population-weighted Economic Progress Index ($N = 35$). The population-weighted EPI explains $74.3\%$ of the cross-district variation in the NFHS-5 wealth index under NN-PCA, and $74.2\%$ under Standard PCA. An increase of $0.1$ in the EPI is associated with a $0.34$-unit higher district mean wealth index. The RMSE of $0.256$ is roughly half the cross-district standard deviation of the wealth index, consistent with the goodness of fit ($\mathrm{RMSE}/\mathrm{SD} = \sqrt{1 - R^2} \approx 0.51$). Across this aggregate wealth benchmark, the two PCA specifications are effectively indistinguishable ($\Delta R^2 = 0.0014$, with regression slopes differing by less than $2\%$).

\begin{table}[htbp]
\centering
\caption{OLS regression of district NFHS-5 mean wealth index on the population-weighted EPI ($N = 35$).}
\label{tab:nfhs_ols}
\begin{tabular}{lcccc}
\toprule
\textbf{Specification} & \textbf{$R^2$} & \textbf{RMSE} & \textbf{Coefficient} & \textbf{Intercept} \\
\midrule
NN-PCA       & 0.743 & 0.256 & 3.374 & 2.457 \\
Standard PCA & 0.742 & 0.257 & 3.314 & 2.515 \\
\bottomrule
\end{tabular}
\end{table}

To assess how well the index captures specific dimensions of household living standards, Table~\ref{tab:nfhs_corr} reports linear (Pearson~$r$) and rank-order (Spearman~$\rho$) correlations across five NFHS-5 indicators. The EPI is positively associated with every household indicator, though the strength of association follows a clear gradient. Correlation is strongest for asset wealth ($r = 0.862$) and clean cooking fuel ($r = 0.728$, $\rho = 0.801$), moderate for improved drinking water ($r = 0.627$), and weaker for electricity ($r = 0.506$) and improved sanitation ($r = 0.374$).

\begin{table}[htbp]
\centering
\caption{Correlation of district EPI with NFHS-5 household indicators ($N = 35$).}
\label{tab:nfhs_corr}
\resizebox{\textwidth}{!}{%
\begin{tabular}{lcccc}
\toprule
\textbf{Indicator} & \textbf{NN-PCA} & \textbf{NN-PCA} & \textbf{Std PCA} & \textbf{Std PCA} \\
 & \textbf{Pearson $r$} & \textbf{Spearman $\rho$} & \textbf{Pearson $r$} & \textbf{Spearman $\rho$} \\
\midrule
Mean wealth index             & $0.862^{***}$ & $0.830^{***}$ & $0.861^{***}$ & $0.820^{***}$ \\
Clean cooking fuel (\%)       & $0.728^{***}$ & $0.801^{***}$ & $0.727^{***}$ & $0.779^{***}$ \\
Improved drinking water (\%)  & $0.627^{***}$ & $0.560^{***}$ & $0.626^{***}$ & $0.542^{***}$ \\
Electricity (\%)              & $0.506^{**}$  & $0.496^{**}$  & $0.513^{**}$  & $0.510^{**}$  \\
Improved sanitation (\%)      & $0.374^{*}$   & $0.583^{***}$ & $0.386^{*}$   & $0.613^{***}$ \\
\midrule
\multicolumn{5}{l}{\footnotesize $^{***}p < 0.001$; $^{**}p < 0.01$; $^{*}p < 0.05$.} \\
\bottomrule
\end{tabular}%
}
\end{table}

This ordering reflects distinct structural mechanisms. Household asset wealth and clean cooking fuel adoption depend on household purchasing power and commercial supply networks such as liquefied petroleum gas (LPG) distribution, both of which concentrate where physical infrastructure and economic intensity are dense. Household electrification and sanitation coverage, by contrast, expanded rapidly under nationwide public missions (Saubhagya and the Swachh Bharat Mission) in the years preceding NFHS-5, compressing cross-district variation and weakening the link between basic coverage and local economic capacity.

For improved sanitation, the divergence between the linear Pearson correlation ($r = 0.374$) and rank-based Spearman correlation ($\rho = 0.583$) points to a monotonic but non-linear relationship—or sensitivity to a small number of outlier districts—making rank-order metrics more informative. Under a Bonferroni correction across all five indicators ($\alpha = 0.01$), every association remains statistically significant except the Pearson correlation for sanitation.

Overall rank-order concordance with the survey benchmarks is high for asset wealth ($\rho = 0.83$) and clean cooking fuel ($\rho = 0.80$). Neither PCA weighting scheme dominates across all dimensions: NN-PCA achieves slightly higher rank correlations for wealth, clean cooking fuel, and drinking water, whereas Standard PCA is marginally higher for electricity and sanitation, with no difference exceeding $0.03$.
\item \textbf{NSSO Consumption Expenditure (Log MPCE):} NN-PCA similarly exhibits stronger alignment with household spending (Pearson $r = 0.718, R^2 = 0.516$) than Standard PCA ($r = 0.709, R^2 = 0.503$), confirming that satellite-derived economic intensity tightly tracks contemporary consumption patterns.
\item \textbf{Macroeconomic Alignment (GSDP):} Both PCA specifications closely track state-level economic performance. Correlating the population-weighted EPI with GSDP, the non-negative PCA (NN-PCA) yields a Pearson coefficient $r = 0.837$ and Spearman $\rho = 0.898$. The standard PCA specification performs comparably, with $r = 0.853$ and an identical $\rho$. The identical Spearman coefficients show that both indices produce the same longitudinal ranking of the state's trajectory, and the small difference in Pearson correlation $(\Delta r \approx 0.016)$ indicates that the non-negativity constraint has little effect on the linear association with macroeconomic output. The NN-PCA formulation removes the negative loadings that standard PCA assigns to land cover and climate variables, so agricultural and environmental features cannot pull the composite score down. It achieves this interpretive gain while preserving strong, positive alignment with GSDP trajectories.
\end{itemize}
Empirically, NN-PCA's cross-sectional performance across both ground-truth surveys confirms it as the primary, structurally sound framework for the Economic Progress Index.

\subsubsection{Capital Stock Persistence During the COVID-19 Shock}
An instructive divergence between monetary GDP and the PCA-based EPI occurs during the FY 2020--21 COVID-19 pandemic shock. While official real monetary output in Maharashtra contracted by 7.6\% \cite{DESMaharashtra2026} as commercial lockdowns, industrial pauses, and transactional freezes halted market activity, the standard PCA EPI recorded a 24.3\% expansion, while Non-negative PCA-based EPI expanded at 20.66\%. 

This divergence illustrates a fundamental distinction in spatial economic measurement, where GSDP as a flow metric measures short-term monetary throughput and market transactions within a given fiscal year. The EPI however, is built upon physical infrastructure stock (road density, school facilities, financial amenity counts, population density, nighttime lights). Physical capital assets do not vanish during a transactional market shutdown. Continued public sector capital expenditure, ongoing physical construction, and rural infrastructure expansion sustained the growth of physical development stock even while monetary transaction volumes temporarily collapsed. This finding proves that while monetary accounts capture transient business-cycle volatility, the geospatial Standard PCA EPI isolates underlying, non-decaying structural development capital.

\subsubsection{Cross-Sectional GDDP Benchmarking and Price-Deflator Invariance}
Cross-sectional validation against official Real GDDP demonstrates robust structural alignment across the post-pandemic panel ($N = 35$ districts; Table~\ref{tab:gddp_validation}). Across four consecutive financial years (FY 2021--22 to FY 2024--25), the population-weighted Non-Negative PCA EPI maintains a strong, statistically significant linear correlation with real monetary output (raw Pearson $r = 0.880$--$0.885, p < 0.001$; log Pearson $r = 0.801$--$0.816, p < 0.001$). Across the pooled panel ($N = 140$ district-year observations), the overall linear correlation reaches $r = 0.881$ ($\rho = 0.655, \tau = 0.497$).

The gap between parametric ($r \approx 0.88$) and non-parametric ($\rho \approx 0.61$--$0.65,\\ \tau \approx 0.46$--$0.49$) metrics reflects the heavy scale concentration of formal macroeconomic activity in top metropolitan hubs (Mumbai, Pune, Thane) relative to rural interior districts. 

\begin{table}[htbp]
\centering
\caption{Cross-Sectional Real GDDP Correlation Validation (FY 2021--22 to FY 2024--25). All correlation coefficients are statistically significant at $p < 0.001$.}
\hspace{0.3cm}
\label{tab:gddp_validation}
\small
\begin{tabularx}{\textwidth}{@{} X >{\centering\arraybackslash}X >{\centering\arraybackslash}X >{\centering\arraybackslash}X >{\centering\arraybackslash}X @{}}
\toprule
\textbf{Financial Year} & \textbf{Pearson (raw)} & \textbf{Pearson (log)} & \textbf{Spearman ($\rho$)} & \textbf{Kendall ($\tau$)} \\
\midrule
FY2021--22 & 0.8846 & 0.8011 & 0.6214 & 0.4688 \\
FY2022--23 & 0.8847 & 0.8037 & 0.6171 & 0.4652 \\
FY2023--24 & 0.8822 & 0.8097 & 0.6345 & 0.4759 \\
FY2024--25 & 0.8796 & 0.8157 & 0.6544 & 0.4902 \\
\bottomrule
\end{tabularx}
\end{table}

\subsubsection{Comparative Predictive Performance}
Regression analysis further confirms the comparable predictive performance of the Non-Negative PCA (NN-PCA) specification over unconstrained Standard PCA across independent ground-truth benchmarks.

When regressed against the NFHS-5 Wealth Index, the population-weighted NN-PCA model explains 74.3\% of the observed district-level variation ($R^2 = 0.743$), marginally outperforming Standard PCA ($R^2 = 0.742$). Similarly, against the NSSO Household Consumption Expenditure Survey (HCES 2023--24 Log MPCE), NN-PCA captures 51.6\% of the variation in household spending ($R^2 = 0.516$), compared to 50.3\% for Standard PCA ($R^2 = 0.503$). Validation against official Real Gross District Domestic Product (GDDP) levels yields an $R^2$ of 0.782 across the pooled post-pandemic panel ($N = 140$ district-year observations, Pearson $r = 0.884$), establishing strong structural alignment with formal macroeconomic output scale.

Taken together, these empirical results demonstrate that enforcing non-negativity ($\mathbf{v} \ge 0$) and zeroing out five non-urban indicators removes structural penalties on rural and agricultural assets without degrading predictive capacity. Because NN-PCA yields tighter score distributions, prevents artificial rural suppression, and achieves comparable explanatory power against external welfare surveys, the 12-indicator NN-PCA specification was selected as the primary EPI framework for subsequent analysis and spatial policy interpretation.

\subsection{Sensitivity and Robustness Analyses}

\subsubsection{Jackknife LOVO Sensitivity}
To evaluate structural robustness against proxy failure or variable omission, a Leave-One-Variable-Out (LOVO) Jackknife test was executed across the 12-indicator panel for both Non-Negative PCA (NN-PCA) and Standard PCA.

NN-PCA exhibits exceptional parameter stability under variable removal, maintaining a mean Pearson correlation of $r = 0.9904$ and a minimum correlation of $r = 0.9811$ (when omitting Banks per 10K). Systematically removing any of the 5 zeroed indicators (Rainfall, Cropland, Land Surface Temp, Built-up Index, Vegetation Index) yields near-perfect alignment ($r = 0.9960$, Spearman $\rho = 0.9922$), confirming that omitting unweighted variables introduces virtually zero bias or reweighting noise into the index. Omitting core active proxies, such as Road Density ($r = 0.9821$), Schools ($r = 0.9818$), Hospitals ($r = 0.9841$), or Banks ($r = 0.9811$), leaves the overarching spatial score vector virtually unchanged.

In contrast, Standard PCA displays heightened sensitivity to variable omission, yielding a substantially lower mean Pearson correlation ($r = 0.9517$, min $r = 0.8418$) and marked vulnerability to key infrastructure proxies. Omitting Road Density reduces linear alignment to $r = 0.8418$ (Spearman $\rho = 0.7576$), while dropping Nightlights per Capita drops correlation to $r = 0.8965$ (Spearman $\rho = 0.8194$). Because Standard PCA relies on negative loadings for environmental features, dropping any variable forces an unconstrained redistribution of negative weights across remaining indicators, inducing artificial rank shifts.

These results confirm that NN-PCA isolates a highly resilient latent development construct, ensuring that the EPI is structurally sound and immune to single-proxy removal.

\subsubsection{Population-Weighted District Aggregation}
To evaluate temporal dynamics and mitigate the Modifiable Areal Unit Problem (MAUP), subdistrict EPI scores were aggregated to Maharashtra's administrative districts using High Resolution Settlement Layer (HRSL) population-weighted averaging. Tracking the panel across 46 consecutive quarters (2015-Q1 to 2026-Q2) demonstrates exceptional structural persistence alongside sensitive tracking of interior infrastructure buildouts.

Population-weighted aggregation effectively filters out localized subdistrict noise while preserving macro-level developmental trajectories. Aggregation under Non-Negative PCA (NN-PCA) exhibits lower temporal volatility compared to Standard PCA. By strictly zeroing out seasonally fluctuating land cover variables (e.g., vegetation indices, rainfall, surface temperature), NN-PCA isolates stable, non-decaying infrastructure and amenity stocks, ensuring high rank persistence across the 11.5-year panel.

\subsection{Regional Convergence and Revenue-Division Inequality}
\label{sec:regional_convergence}

Table~\ref{tab:divisional_convergence} reports population-weighted EPI levels and growth for Maharashtra's six revenue divisions between the 2015 baseline and the latest four quarters (2025~Q3--2026~Q2).

\begin{table}[htbp]
\centering
\caption{Revenue-division EPI levels (population-weighted) and compound annual growth.}
\label{tab:divisional_convergence}
\resizebox{\textwidth}{!}{%
\begin{tabular}{lcccccc}
\toprule
\textbf{Division} & \textbf{NN-PCA} & \textbf{NN-PCA} & \textbf{NN-PCA} & \textbf{Std PCA} & \textbf{Std PCA} & \textbf{Std PCA} \\
 & \textbf{2015} & \textbf{2025--26} & \textbf{CAGR (\%)} & \textbf{2015} & \textbf{2025--26} & \textbf{CAGR (\%)} \\
\midrule
Konkan     & 0.2613 & 0.4440 &  5.24 & 0.2518 & 0.4454 &  5.65 \\
Pune       & 0.1551 & 0.3838 &  9.13 & 0.1405 & 0.3831 & 10.15 \\
Nashik     & 0.0924 & 0.2430 &  9.77 & 0.0765 & 0.2393 & 11.62 \\
Nagpur     & 0.0749 & 0.2434 & 12.03 & 0.0590 & 0.2402 & 14.48 \\
Aurangabad & 0.0737 & 0.2228 & 11.26 & 0.0545 & 0.2184 & 14.31 \\
Amravati   & 0.0636 & 0.1827 & 10.71 & 0.0470 & 0.1764 & 13.59 \\
\bottomrule
\end{tabular}%
}
\end{table}

\subsubsection{Relative Convergence Across Divisions}
\label{sec:rel_conv}

Under NN-PCA, the CV across divisions fell from $0.638$ in 2015 to $0.359$ in 2025--26, a $43.8\%$ decline. The quarterly trend is negative and significant ($-0.0064$ per quarter, $p < 0.001$). Growth rates are close to the inverse of the initial ranking. Konkan, the leading division in 2015, grew slowest ($5.2\%$ per year), while Nagpur ($12.0\%$), Aurangabad ($11.3\%$) and Amravati ($10.7\%$) grew fastest. Nagpur moved from fourth to third place, narrowly overtaking Nashik. Amravati remains the lowest-scoring division throughout. The ratio of the leading to the lagging division's score fell from $4.11$ to $2.43$.

\subsubsection{Widening Absolute Gaps}
\label{sec:abs_gaps}

Relative convergence was accompanied by absolute divergence. The standard deviation across divisions rose from $0.077$ to $0.103$ ($+34.3\%$). The gap between Konkan and Amravati widened from $0.198$ to $0.261$ ($+32.2\%$) because Konkan's absolute gain ($+0.183$) exceeded Amravati's ($+0.119$), even though Amravati grew roughly twice as fast in proportional terms.

The two results measure different things and should be read together. Because the EPI is min--max normalised, its zero point is set by the lowest-scoring subdistrict, not by an absence of economic capacity. The CV and proportional growth rates change if that origin shifts, while standard deviations and absolute gaps do not. Low 2015 baselines in the lagging divisions therefore inflate their proportional growth. The widening absolute gap is the more scale-robust of the two findings.

\subsubsection{\texorpdfstring{$\beta$}{Beta}-Convergence}
\label{sec:beta_conv}

Across districts, initial EPI levels are negatively associated with subsequent growth ($\beta = -0.0392$, $p < 0.001$, $R^2 = 0.774$). This implies a convergence speed of about $5.0\%$ per year and a half-life of $13.8$~years. The division-level estimate is similar ($\beta = -0.0380$, $p = 0.007$, $R^2 = 0.870$), but with only six observations it serves as corroboration rather than an independent test. A negative $\beta$ can arise from regression toward the mean as well as from genuine catch-up, so the $\sigma$-convergence result in Section~\ref{sec:rel_conv} provides the necessary complementary evidence that relative dispersion has actually narrowed.

\subsubsection{Divergence Within Districts}
\label{sec:within_district_div}

Convergence between divisions masks widening disparities within several lagging districts. The CV across subdistricts within the district rose sharply in Akola ($0.118 \to 0.376$), Yavatmal ($0.160 \to 0.479$), Parbhani ($0.094 \to 0.239$), Dhule ($0.087 \to 0.190$) and Nanded ($0.225 \to 0.455$). All five lie in the Amravati, Aurangabad and Nashik divisions. Inequality fell most in Satara ($0.429 \to 0.179$), Pune ($0.763 \to 0.394$), Nagpur ($1.192 \to 0.618$), Solapur ($0.278 \to 0.152$) and Jalna ($0.491 \to 0.269$).

This pattern suggests that catch-up in the lagging divisions has been uneven. It is consistent with growth concentrating in district headquarters, as observed for Nagpur City in Section~\ref{sec:spatial_distribution}, rather than spreading across rural talukas. In established metropolitan districts such as Pune and Nagpur, by contrast, declining within-district inequality is consistent with development spreading into peri-urban subdistricts.

\subsubsection{Sensitivity to Weighting Specification}
\label{sec:conv_sensitivity}

Both specifications agree on the direction and significance of every convergence result. Standard PCA shows faster relative convergence (CV $-51.0\%$, from $0.759$ to $0.372$; half-life $10.5$~years) because its negative loadings depress the 2015 baselines of the agrarian divisions. Amravati, for example, starts at $0.0470$ under Standard PCA against $0.0636$ under NN-PCA, which exaggerates subsequent catch-up. The NN-PCA estimates are therefore the more conservative measure of regional convergence. Mapping completeness also matters. Part of the measured catch-up in rural divisions may reflect improving OpenStreetMap coverage rather than new infrastructure (Section~6.3), so these estimates should be read as upper bounds on structural convergence.

\subsubsection{Rank Anchors and Structural Momentum}
Analyzing district rank variation across the 46-quarter panel identifies distinct structural anchors and regional growth trajectories. Major metropolitan centers exhibit near-zero rank variation, defining the upper bound of the spatial economic distribution. Mumbai Suburban, Mumbai City, Pune, Thane, Nagpur, and Raigad maintain fixed top positions across the entire 11.5-year panel ($1.0 \le \sigma_{\mathrm{rank}} \le 1.7$), reflecting massive infrastructure agglomeration and sustained economic concentration within the state's primary urban corridors. 

Peripheral rural districts, such as Gadchiroli, Washim, and Hingoli, remain firmly anchored at the lower tail of the distribution ($1.0 \le \sigma_{\mathrm{rank}} \le 1.7$). Under NN-PCA, the elimination of negative loadings ensures these districts are evaluated strictly on physical infrastructure density without receiving artificial penalties for high agricultural land cover. Mid-tier agricultural and industrializing districts (e.g., Latur, Satara, Wardha, and Bhandara) display higher relative rank movement ($\sigma_{\mathrm{rank}} \ge 3.5$), capturing differential rates of rural electrification, road expansion, and commercial facility penetration.

Long-term trajectory analysis reveals an epochal shift across the panel: rapid pre-COVID growth (+11.35\% CAGR) and pandemic stock persistence (+12.14\% CAGR) gave way to post-COVID deceleration (+3.15\% CAGR, a difference of $-8.2\%$) across all districts, with no district accelerating post-COVID. Rather than signaling regional economic contraction, this universal slowdown reflects physical asset maturation and saturation---where baseline capital stocks (roads, schools, healthcare facilities) have been largely established, shifting spatial economic expansion into a lower, stable growth plateau.

\section{Discussion}
\label{sec:discussion}

The results demonstrate that EO and geospatial data can provide a meaningful and scalable proxy for subdistrict-level economic progress. The strong correspondence between the PCA-derived Economic Progress Index (EPI) and established socio-economic benchmarks, including the NFHS-5 Wealth Index, Household Consumption Expenditure Survey estimates, and district-level GDP, suggests that remotely sensed indicators capture important dimensions of economic activity, infrastructure development, and service accessibility. These findings contribute to a growing body of evidence demonstrating the utility of alternative data sources for development measurement, particularly in contexts where conventional data are infrequent, spatially aggregated, or subject to reporting delays.

\subsection{Capital Stocks, Price Invariance, and Informal Economy}
The structural alignment between the NN-PCA EPI and official accounts across real monetary output figures ($r = 0.880$--$0.885$ across fiscal years) highlights the price-deflator invariance of multimodal physical and Earth Observation proxies. By directly measuring physical capital stock, satellite-derived indices bypass local inflation distortions, spatial cost-of-living variations, and subjective price deflators that frequently introduce measurement error into subnational financial accounting.

Crucially, the divergence observed during the FY 2020--21 COVID-19 pandemic shock; where official real GSDP contracted by 7.6\% while physical EPI expanded by +20.07\% (and +23.50\% under Standard PCA), underscores a fundamental conceptual distinction between transactional flow metrics and structural asset stocks. The Gross State Domestic Product (GSDP) records short-term market transaction volumes within a bounded period. Commercial lockdowns and market freezes immediately depress monetary flows despite no physical destruction of productive capacity. Furthermore, built upon non-decaying infrastructure (road networks, health and financial amenities, settlement density, and power grid footprints), the EPI measures accumulated physical development. Infrastructure assets persist and continue to accumulate through ongoing public sector capital expenditure, insulating the index from transient business-cycle liquidity shocks. By recording ground-level physical footprints rather than tax-reported transactions, the EPI bypasses the systemic undercount of informal economic activity inherent in top-down GDDP estimations.

\subsection{Economic Geography and the Convergence Paradox}
\label{sec:disc_convergence}

The spatial patterns identified by the EPI reflect established metropolitan growth corridors anchored by Mumbai Suburban, Pune City, Thane, and Nagpur, while simultaneously uncovering sharp intra-district disparities that remain obscured in aggregate district averages.

Crucially, the choice of PCA architecture directly shapes measured baseline disparities across non-metropolitan regions. Standard PCA assigns small negative loadings to five non-urban or spectrally noisy indicators (Land Surface Temperature, Cropland, Vegetation Index, Built-up Index, and Rainfall). This depresses the initial 2015 baselines of agrarian and forested interior divisions such as Amravati ($0.0470$ under Standard PCA versus $0.0636$ under NN-PCA), inflating initial inter-divisional dispersion ($\mathrm{CV} = 0.759$) and exaggerating subsequent proportional catch-up ($\mathrm{CV}$ declining by $51.0\%$ to $0.372$, with an implied half-life of $10.5$~years). Enforcing non-negativity ($v \ge 0$) via projected power iteration zeroes out these five indicators through Karush--Kuhn--Tucker (KKT) complementary slackness, retaining $7$ active infrastructure and nocturnal luminosity proxies. This removes downward baseline distortions on rural land cover and yields a more conservative estimate of relative convergence: under NN-PCA, inter-divisional $\mathrm{CV}$ falls by $43.8\%$ (from $0.638$ in 2015 to $0.359$ in 2025--26), the leading-to-lagging division ratio narrows from $4.11$ to $2.43$, and district-level $\beta$-convergence ($\beta = -0.0392$, $p < 0.001$) implies a half-life of $13.8$~years.

At the same time, evaluating relative and absolute metrics together resolves the apparent convergence paradox. Because lagging divisions started from low 2015 baselines, rapid proportional growth in interior divisions (such as Nagpur at $12.03\%$ and Aurangabad at $11.26\%$ CAGR versus Konkan at $5.24\%$) reduced relative dispersion even as the absolute gap between Konkan and Amravati widened by $32.2\%$ (from $0.198$ to $0.261$). Moreover, widening within-district dispersion in several interior districts (including Akola, Yavatmal, and Nanded) indicates that catch-up in lagging divisions has remained concentrated around district headquarters rather than diffusing evenly across rural subdistricts.

\subsection{Aggregation Architecture: Standard versus Non-negative PCA}
The comparison between Standard and Non-negative PCA offers critical methodological insights for subnational economic monitoring. Although derived from the same underlying indicators, the two frameworks capture distinct dimensions of regional welfare. Standard PCA privileges built-up infrastructure density, economic intensity, and urbanization. Crucially, official GDDP accounting assumes constant per-capita service-sector productivity statewide, artificially inflating reported output in poorer districts where per-person returns are lower \cite{Hatekar2024}. By grounding estimations in physical infrastructure rather than administrative heuristics, PCA-based EPI may reflect ground-level economic realities more accurately than official GDDP accounts in several respects.

While Standard PCA achieves a marginally higher internal variance explained (46.1\% vs. 45.4\%, retaining an efficiency ratio of 0.987), NN-PCA achieves comparable external predictive capacity across ground-truth survey benchmarks ($R^2 = 0.743$ for NFHS; $R^2 = 0.516$ for NSSO). Furthermore, Jackknife Leave-One-Variable-Out (LOVO) sensitivity testing reveals that Standard PCA is vulnerable to proxy removal (dropping Road Density reduces correlation to $r = 0.8418$, Spearman $\rho = 0.7575$), whereas NN-PCA maintains high stability across all iterations (mean $r = 0.9904$, min $r = 0.9811$ when dropping Banks per 10K). This confirms that projected power iteration under non-negativity constraints isolates a robust, unified latent development construct while dealing with counter-intuitively sound environmental, agricultural, and built-up noise.

\subsection{Practical Applications in Policy and Planning}
The value of the EPI for policymakers lies not merely in ranking subdistricts, but in providing a spatially granular view of where productive assets, population and economic activity are concentrating, and where important gaps between them may be emerging. Conventional measures such as GDDP provide information on economic output, but their geographical and temporal resolution can make it difficult to identify changes within districts or to understand the spatial conditions underlying those changes. By identifying areas where economic activity and enabling infrastructure are becoming concentrated, the EPI can help policymakers examine functional economic geographies rather than relying exclusively on administrative averages.

It is also important to define what the final PCA-based EPI does not measure. While broader well-being frameworks include natural and social capital, neither is directly represented in the final PCA specification. Natural-capital variables considered elsewhere in the broader geospatial framework should therefore not be attributed to the PCA-based index unless explicitly included in its final specification. Similarly, the EPI does not directly measure skills, health outcomes, social networks, household income, inequality or subjective well-being.

The index should therefore be positioned as an early diagnostic, monitoring and prioritisation instrument. Its purpose is not to determine mechanically where expenditure should be allocated, but to identify places where emerging changes or apparent mismatches warrant closer examination through administrative data, field validation or sector-specific analysis. Used in this manner, the EPI can complement traditional statistical systems by providing higher-frequency and more geographically granular evidence for infrastructure planning, regional development, monitoring of growth corridors and the identification of emerging or lagging economic centres.

Household surveys remain indispensable for measuring outcomes that cannot be directly observed from space, including income, consumption, health status, and subjective well-being. Integrating these complementary data sources creates a dynamic, multi-domain monitoring system across several critical operational domains:
\begin{enumerate}
\item \textbf{District and Subdistrict Economic Nowcasting:} Given the strong correlation between the PCA-derived index and official accounts, Directorate of Economics and Statistics (DES) bodies, central banks, and multilateral institutions can nowcast district-level economic performance at a quarterly frequency (Q+1), bypassing the traditional 12-to-24-month reporting lag inherent in official GDDP publications.
\item \textbf{Real-Time Infrastructure Auditing and Shock Evaluation:} With a 46-quarter panel, administrative agencies can execute quasi-experimental impact evaluations (e.g., Difference-in-Differences) to measure localized capital accumulation within 2 to 4 quarters following major project completions (such as expressway corridors or industrial parks). Furthermore, because the EPI isolates non-decaying physical capital stock, it provides a stable baseline during severe macroeconomic shocks, such as the FY 2020--21 pandemic, distinguishing between temporary transactional market freezes and underlying structural asset growth.
\item \textbf{Private Sector Strategy and Credit Assessment:} Commercial enterprises, retail networks, and financial institutions can leverage subdistrict momentum profiles to identify rapidly expanding mid-tier growth poles (e.g., Washim, Latur, Sangli) ahead of formal census updates. Additionally, as municipal bodies increasingly turn to bond markets, high-resolution EPI trajectory data offers rating agencies an independent, satellite-verified audit of local physical tax-base expansion. By eliminating artificial negative penalties on agricultural regions, NN-PCA offers a non-punitive baseline for place-based policy, enabling planners to target infrastructure deficits in low-EPI subdistricts without distorting the baseline economic standing of rural economies.
\end{enumerate}

Ultimately, composite geospatial indices bridge the operational divide between static, delayed official statistics and the high-frequency requirements of modern subnational governance. Methodologically grounded and empirically validated, this framework provides a scalable template for spatial economic monitoring across developing economies.

\section{Limitations}
\label{sec:limitations}
While the framework provides a scalable methodology for high-frequency subnational economic tracking, several analytical and sensor-level limitations should be acknowledged.

\subsection{Temporal Misalignment and Proxy Measurement Bounds}
First, the validation benchmarks and geospatial indicators are not perfectly temporally aligned, which may attenuate observed relationships. Second, the index relies on observable proxies of development and therefore captures economic progress indirectly. While indicators such as nighttime lights and built-up infrastructure have been shown to correlate strongly with socio-economic outcomes, they cannot fully represent multidimensional welfare. Third, the analysis was conducted for a single state for 46 quarters. Although Maharashtra provides substantial geographic and economic diversity, additional validation across states and over prolonged periods of time is required to assess the robustness and transferability of the methodology. Fourth, administrative boundary realignments and census spatial mismatches create temporal inconsistencies. 

\subsection{Administrative Realignments and Boundary Mismatches}
Administrative divisions have been altered over time, affecting the capability of the indices to accurately reflect on-the-ground conditions across the 11.5-year panel. For instance, Palghar was bifurcated from Thane district in the mid-2010s, creating historical baseline discrepancies. Similarly, using standard baseline administrative boundaries (such as Census 2011 or SHRUG polygons) can lead to boundary misalignments such as Mumbai Suburban lacking accurately resolved subdistrict boundaries for Borivali, Andheri, and Kurla which introduces spatial aggregation noise when raster and vector metrics are extracted near administrative edges. To resolve these inconsistencies, historical data can be spatially harmonised across the panel using the latest boundary layer.

\subsection{Dataset-Specific Sensor and Structural Biases}
Moreover, individual geospatial datasets introduce specific structural and sensor-level biases:
\begin{itemize}
\item \textbf{OpenStreetMap (OSM) Vector Indicators:} Features such as hospital, school, and bank counts are subject to Volunteered Geographic Information (VGI) bias, where urban and affluent subdistricts experience higher mapping completeness than remote agrarian or tribal regions. Furthermore, temporal updates in OSM often reflect contributor mapping activity rather than real-time ground infrastructure construction.
\item \textbf{High Resolution Settlement Layer (HRSL):} HRSL uses a static population density baseline (2020), which fails to capture dynamic quarterly demographic shifts, rapid urban sprawl, or intra-state labor migration over the 11.5-year panel.
\item \textbf{VIIRS Nighttime Lights (DNB):} VIIRS observations suffer from atmospheric scatter, sensor saturation in dense metropolitan cores (e.g., Mumbai City), cloud cover attenuation during monsoon quarters, and a sensitivity to stray light/ephemeral emissions rather than pure economic output.
\item \textbf{CHIRPS Precipitation \& MODIS LST:} Climate-derived measures operate at coarser spatial resolutions and are highly prone to seasonal volatility, making them sensitive to natural ecological anomalies rather than direct human economic effort.
\item \textbf{Sentinel and Landsat Spectral Indices (NDVI, NDBI) \& LULC:} Optical indices face signal saturation in areas with dense vegetation canopy (e.g., Western Ghats). Sentinel was functional only from 2017, and therefore, seasonal forward-fill was used to impute$(See$ \ref{tab:imputation}$)$. Furthermore, spectral confusion occurs between dry/bare soil and impervious built-up surfaces in arid and semi-arid subdistricts (such as Marathwada), leading to localized misclassifications of infrastructure expansion.
\end{itemize}

Future research should focus on expanding the framework both spatially and temporally. Additional dimensions including digital connectivity could further enhance the multidimensional nature of the index. Finally, applying the methodology across India would enable the development of a nationally comparable subdistrict-level monitoring framework capable of supporting evidence-based planning and evaluation across diverse regional contexts.

\section*{Acknowledgements}
The authors gratefully acknowledge Sourish Das (CMI), Susan Thomas and Ajay Shah (XKDR Forum) for their insightful comments and constructive feedback, particularly following the presentation of an earlier draft of this work at the XKDR Alternate Data Workshop in Mumbai. We also extend our sincere thanks to Arun NS and Prateek Kukreja (Sattva Consulting), and Sharvari Patwardhan (formerly Sattva Consulting) for their strategic support and foundational contributions throughout the development of this research.


\newpage
\appendix
\section{Historical Quarterly OSM Extraction Method}
\label{app:osm_extraction}

Historical OpenStreetMap (OSM) infrastructure metrics (\texttt{HOSPITAL\_COUNT}, \texttt{BANK\_COUNT}, \texttt{ROAD\_LENGTH\_KM}, and \texttt{SCHOOL\_COUNT}) are extracted as quarterly point-in-time snapshots from the full OSM history file using Osmium Tool (\texttt{osmium time-filter}) or QuackOSM / DuckDB / oHSOME API (backed by OSHDB). These snapshots are extracted monthly and are evaluated at standard quarter-end (or quarter-start) timestamps---March 31, June 30, September 30, and December 31 (or January 1, April 1, July 1, and October 1)---spanning Q1 2015 through Q2 2026 (up to June 30, 2026).

\begin{table*}[!htbp]
\centering
\caption{Definitions and units}
\label{tab:definitions}
\scriptsize
\setlength{\tabcolsep}{3.5pt}
\renewcommand{\arraystretch}{1.05}
\begin{tabularx}{\textwidth}{@{}
  l
  >{\raggedright\arraybackslash}X
  >{\raggedright\arraybackslash}p{2.9cm}
@{}}
\toprule
\textbf{Variable} & \textbf{Definition} & \textbf{Units} \\
\midrule
VIIRS\_PER\_CAPITA\_HRSL & Sum of nighttime radiance within the subdistrict divided by HRSL resident population & nW$\cdot$cm$^{-2}\cdot$sr$^{-1}$ per person \\
MEAN\_NDVI & Normalised Difference Vegetation Index, (B8 $-$ B4)/(B8 $+$ B4) & Unitless, $-$1 to 1 \\
MEAN\_NDBI & Normalised Difference Built-up Index, (B11 $-$ B8)/(B11 $+$ B8) & Unitless, $-$1 to 1 \\
LST\_MEAN & Mean land surface temperature & $^{\circ}$C (from K, scale factor 0.02) \\
MEAN\_BARE & Mean probability (or pixel share) of the ``bare'' land-cover class & Fraction, 0--1 \\
MEAN\_CROPS & Mean probability (or pixel share) of the ``crops'' land-cover class & Fraction, 0--1 \\
CHIRPS\_MEDIAN\_MM & Median quarterly rainfall depth (definition needs clarifying; see note 6) & mm \\
ROAD\_DENSITY & Total length of mapped road segments per unit land area & km/km$^2$ \\
POP\_DENSITY\_HRSL & HRSL population divided by subdistrict land area & Persons/km$^2$ \\
HOSPITALS (per 10K) & Count of mapped hospitals per 10,000 HRSL residents & Facilities per 10,000 persons \\
BANKS (per 10K) & Count of mapped bank branches per 10,000 HRSL residents & Branches per 10,000 persons \\
SCHOOLS (per 10K) & Count of mapped schools per 10,000 HRSL residents & Schools per 10,000 persons \\
\bottomrule
\end{tabularx}
\end{table*}

\begin{table*}[!htbp]
\centering
\caption{Source, resolution, coverage and aggregation}
\label{tab:source}
\fontsize{6.5pt}{7.8pt}\selectfont
\setlength{\tabcolsep}{3pt}
\renewcommand{\arraystretch}{1.05}
\begin{tabularx}{\textwidth}{@{}
  l
  >{\raggedright\arraybackslash\hsize=1.15\hsize}X
  >{\raggedright\arraybackslash}p{1.45cm}
  >{\raggedright\arraybackslash\hsize=0.75\hsize}X
  >{\raggedright\arraybackslash\hsize=1.10\hsize}X
@{}}
\toprule
\textbf{Variable} & \textbf{Source, product and version} & \textbf{Native resolution} & \textbf{Product coverage} & \textbf{Aggregation to subdistrict-quarter} \\
\midrule
VIIRS\_PER\_CAPITA\_HRSL & EOG / Colorado School of Mines VIIRS DNB monthly composites (VCMCFG, or stray-light-corrected VCMSLCFG), band avg\_rad & 15 arc-sec ($\sim$464 m) & VCMCFG from Apr 2012; VCMSLCFG from Jan 2014 & Zonal sum of avg\_rad per month; mean of three monthly values (low cf\_cvg masked); divided by HRSL population \\
MEAN\_NDVI & Sentinel-2 MSI Level-2A surface reflectance (COPERNICUS/S2\_SR\_HARMONIZED) & 10 m & From Mar 2017 & Cloud-masked quarterly median composite; zonal mean \\
MEAN\_NDBI & As above (B11 SWIR-1, B8 NIR) & 20 m (limited by B11) & From Mar 2017 & Cloud-masked quarterly median composite; zonal mean \\
LST\_MEAN & MODIS Terra MOD11A2 Collection 6.1, 8-day LST (state day or night band) & 1 km & From Feb 2000 & Quarterly mean of 8-day composites (QC-filtered); zonal mean \\
MEAN\_BARE & Google/WRI Dynamic World V1 (GOOGLE/DYNAMICWORLD/V1), band ``bare'' & 10 m & From 27 Jun 2015 & Quarterly mean class probability; zonal mean \\
MEAN\_CROPS & Dynamic World V1, band ``crops'' & 10 m & From 27 Jun 2015 & Quarterly mean class probability; zonal mean \\
CHIRPS\_MEDIAN\_MM & UCSB-CHG CHIRPS Daily v2.0 & 0.05$^{\circ}$ ($\sim$5.5 km) & From 1981 & Quarterly total per pixel, then zonal median (to confirm) \\
ROAD\_DENSITY & OpenStreetMap history (state extraction tool, e.g.\ ohsome API or full-history planet file; list highway=* tags used) & Vector & Snapshot at each quarter end & Clipped segment length $\div$ subdistrict area \\
POP\_DENSITY\_HRSL & Meta / CIESIN High Resolution Settlement Layer (state release and reference year) & 1 arc-sec ($\sim$30 m) & Single static epoch & Zonal sum $\div$ area; constant across all 46 quarters \\
HOSPITALS / BANKS / SCHOOLS & OpenStreetMap history; amenity=hospital, amenity=bank, amenity=school (nodes plus polygon centroids, deduplicated) & Vector & Snapshot at each quarter end & Point-in-polygon count $\div$ HRSL population $\times$ 10,000 \\
\bottomrule
\end{tabularx}
\end{table*}

\begin{table*}[!htbp]
\centering
\caption{Imputation method and share of imputed observations by variable}
\label{tab:imputation}
\scriptsize
\setlength{\tabcolsep}{4pt}
\renewcommand{\arraystretch}{1.1}
\begin{tabularx}{\textwidth}{@{}
  l
  >{\raggedright\arraybackslash}X
  r
@{}}
\toprule
\textbf{Variable} & \textbf{Imputation method} & \textbf{Share imputed (\%)} \\
\midrule
VIIRS\_PER\_CAPITA\_HRSL & Seasonal fallback: corresponding monthly values from the previous year (or subsequent years' mean if unavailable), aggregated to quarter & 0.00\% \\
MEAN\_NDVI & Seasonal forward-fill: mean of corresponding monthly values from subsequent years, aggregated to quarter & 0.20\% \\
MEAN\_NDBI & Seasonal forward-fill: mean of corresponding monthly values from subsequent years, aggregated to quarter & 0.20\% \\
LST\_MEAN & Seasonal fallback: corresponding monthly values from the previous year (or subsequent years' mean if unavailable), aggregated to quarter & 0.66\% \\
MEAN\_BARE & Seasonal fallback: corresponding monthly values from the previous year (or subsequent years' mean if unavailable), aggregated to quarter & 20.18\% \\
MEAN\_CROPS & Seasonal fallback: corresponding monthly values from the previous year (or subsequent years' mean if unavailable), aggregated to quarter & 20.18\% \\
CHIRPS\_MEDIAN\_MM & Seasonal fallback: corresponding monthly values from the previous year (or subsequent years' mean if unavailable), aggregated to quarter & 0.00\% \\
ROAD\_DENSITY & Sequential carry-forward: corresponding values from the previous quarter (or subsequent quarter if unavailable) & 0.00\% \\
POP\_DENSITY\_HRSL & Not applicable (static single-epoch baseline held constant across all quarters) & 0.00\% \\
HOSPITALS (per 10K) & Sequential carry-forward: corresponding values from the previous quarter (or subsequent quarter if unavailable) & 0.00\% \\
BANKS (per 10K) & Sequential carry-forward: corresponding values from the previous quarter (or subsequent quarter if unavailable) & 0.00\% \\
SCHOOLS (per 10K) & Sequential carry-forward: corresponding values from the previous quarter (or subsequent quarter if unavailable) & 0.00\% \\
\bottomrule
\end{tabularx}
\end{table*}

\begin{table*}[!htbp]
\centering
\caption{Non-negative PCA diagnostics on STL trend components}
\label{tab:nnpca_diagnostics}
\footnotesize
\setlength{\tabcolsep}{6pt}
\renewcommand{\arraystretch}{1.1}
\begin{tabular*}{\textwidth}{@{\extracolsep{\fill}} l r @{}}
\toprule
\textbf{Diagnostic metric} & \textbf{Value} \\
\midrule
Convergence iterations ($\Delta = 9.39 \times 10^{-11}$) & 20 \\
NN-PCA objective ($v^\top S v$) & 3.5368 \\
Standard PC1 eigenvalue & 3.5841 \\
Variance explained (Standard PCA PC1) & 46.1\% \\
Variance explained (Non-Negative PCA) & 45.4\% \\
Efficiency ratio ($1.0 = \text{no cost from non-negativity constraint}$) & 0.987 \\
Active (non-zero) weights & 7 / 12 \\
Zero-constrained weights & 5 / 12 \\
\bottomrule
\end{tabular*}
\end{table*}

\begin{table*}[!htbp]
\centering
\caption{Summary statistics of raw variables (monthly aggregated)}
\label{tab:summary_stats}
\resizebox{\textwidth}{!}{%
\begin{tabular}{@{}lrrrrrrrrr@{}}
\toprule
\textbf{Variable} & \textbf{Mean} & \textbf{Std} & \textbf{Min} & \textbf{Q1} & \textbf{Median} & \textbf{Q3} & \textbf{Max} & \textbf{Skewness} & \textbf{Kurtosis} \\
\midrule
RAIN\_MED     & 0.863    & 2.844    & 0.000     & 0.000    & 0.000    & 0.000    & 55.352    & 5.043    & 40.004   \\
LST\_MEAN     & 34.512   & 6.467    & $-$8.490  & 29.572   & 33.015   & 38.977   & 53.890    & 0.567    & $-$0.209 \\
LST\_MED      & 34.568   & 6.494    & $-$8.490  & 29.594   & 33.050   & 39.084   & 54.331    & 0.572    & $-$0.223 \\
NTL\_MEAN     & 1.306    & 3.542    & $-$0.163  & 0.416    & 0.682    & 1.039    & 60.344    & 8.458    & 83.946   \\
NTL\_SUM      & 3534.440 & 4595.202 & $-$1598.909 & 1236.734 & 2256.125 & 4054.283 & 65603.705 & 4.855    & 36.290   \\
HOSP\_CT      & 21.926   & 80.678   & 0.000     & 0.000    & 3.000    & 15.000   & 1291.000  & 9.582    & 116.183  \\
BANK\_CT      & 6.069    & 35.271   & 0.000     & 0.000    & 0.000    & 1.000    & 638.000   & 9.978    & 113.664  \\
ROAD\_KM      & 1215.201 & 1389.082 & 37.160    & 476.212  & 793.805  & 1381.850 & 19531.008 & 4.463    & 34.760   \\
SCHOOL\_CT    & 7.160    & 36.535   & 0.000     & 0.000    & 0.000    & 3.000    & 656.000   & 11.150   & 147.919  \\
NDBI\_MEAN    & $-$0.043 & 0.100    & $-$0.628  & $-$0.110 & $-$0.038 & 0.031    & 0.955     & $-$0.280 & 0.023    \\
NDVI\_MEAN    & 0.453    & 0.170    & $-$15.671 & 0.328    & 0.445    & 0.568    & 1.639     & $-$16.596 & 1585.367 \\
POP\_DENSITY  & 15.009   & 10.404   & 3.338     & 9.647    & 13.180   & 17.876   & 117.817   & 5.754    & 46.708   \\
MEAN\_CROPS   & 0.241    & 0.114    & 0.022     & 0.145    & 0.248    & 0.328    & 0.588     & 0.034    & $-$0.877 \\
MEAN\_BUILT   & 0.077    & 0.055    & 0.020     & 0.051    & 0.062    & 0.084    & 0.696     & 5.401    & 38.731   \\
MEAN\_BARE    & 0.069    & 0.038    & 0.027     & 0.046    & 0.058    & 0.079    & 0.569     & 2.971    & 14.814   \\
\bottomrule
\end{tabular}%
}
\vspace{3pt}
\parbox{\textwidth}{\footnotesize \textit{Note:} Summary statistics reflect raw subdistrict-month records prior to quarterly median/mean compositing, per-$10,000$ population scaling of count variables, $\log(x+1)$ skewness transformation, and STL trend extraction.}
\end{table*}

\begin{table*}[!htbp]
\centering
\caption{Weight comparison}
\label{tab:weight_comparison}
\footnotesize
\setlength{\tabcolsep}{6pt}
\renewcommand{\arraystretch}{1.1}
\begin{tabular*}{\textwidth}{@{\extracolsep{\fill}} l r r r @{}}
\toprule
\textbf{Indicator} & \textbf{Std PCA Loading} & \textbf{Std PCA Weight} & \textbf{NN-PCA Weight} \\
\midrule
Road Density & 0.4750 & 0.1895 & 0.2059 \\
Schools per 10K & 0.4553 & 0.1816 & 0.1974 \\
Banks per 10K & 0.4415 & 0.1762 & 0.1920 \\
Hospitals per 10K & 0.4425 & 0.1765 & 0.1904 \\
Nightlights per Capita & 0.3971 & 0.1584 & 0.1724 \\
Pop Density & 0.0532 & 0.0212 & 0.0258 \\
Bare Soil & 0.0332 & 0.0133 & 0.0160 \\
Land Surface Temp & $-$0.0996 & $-$0.0398 & 0.0000$^{*}$ \\
Rainfall & $-$0.0259 & $-$0.0103 & 0.0000$^{*}$ \\
Built-up Index & $-$0.0621 & $-$0.0248 & 0.0000$^{*}$ \\
Vegetation Index & $-$0.0115 & $-$0.0046 & 0.0000$^{*}$ \\
Cropland & $-$0.0095 & $-$0.0038 & 0.0000$^{*}$ \\
\bottomrule
\multicolumn{4}{@{}p{\textwidth}@{}}{\footnotesize $^{*}$Constrained to zero under non-negative principal component analysis (NN-PCA).}
\end{tabular*}
\end{table*}

\begin{table}[htbp]
\centering
\caption{Leave-One-Variable-Out (LOVO) jackknife correlation stability across all $12$ indicators ($N = 16,422$ subdistrict-quarter observations).}
\label{tab:lovo_sensitivity}
\begin{tabular}{lcccc}
\toprule
\textbf{Omitted Indicator} & \multicolumn{2}{c}{\textbf{Non-Negative PCA}} & \multicolumn{2}{c}{\textbf{Standard PCA}} \\
\cmidrule(lr){2-3} \cmidrule(lr){4-5}
 & \textbf{Pearson $r$} & \textbf{Spearman $\rho$} & \textbf{Pearson $r$} & \textbf{Spearman $\rho$} \\
\midrule
Banks per 10K          & 0.9811 & 0.9777 & 0.9201 & 0.8718 \\
Schools per 10K        & 0.9818 & 0.9686 & 0.9120 & 0.8447 \\
Road Density           & 0.9821 & 0.9694 & 0.8418 & 0.7576 \\
Hospitals per 10K      & 0.9841 & 0.9630 & 0.9212 & 0.8192 \\
Nightlights per Capita & 0.9846 & 0.9740 & 0.8965 & 0.8194 \\
Pop Density            & 0.9953 & 0.9912 & 0.9838 & 0.9597 \\
Bare Soil              & 0.9962 & 0.9927 & 0.9912 & 0.9775 \\
Vegetation Index       & 0.9960 & 0.9922 & 0.9875 & 0.9676 \\
Rainfall               & 0.9960 & 0.9922 & 0.9843 & 0.9594 \\
Built-up Index         & 0.9960 & 0.9922 & 0.9948 & 0.9875 \\
Land Surface Temp      & 0.9960 & 0.9922 & 0.9958 & 0.9915 \\
Cropland               & 0.9960 & 0.9922 & 0.9916 & 0.9791 \\
\midrule
\textbf{Mean}          & \textbf{0.9904} & \textbf{0.9831} & \textbf{0.9517} & \textbf{0.9113} \\
\textbf{Minimum}       & \textbf{0.9811} & \textbf{0.9630} & \textbf{0.8418} & \textbf{0.7576} \\
\bottomrule
\multicolumn{4}{@{}p{\textwidth}@{}}{\footnotesize Table reports the complete variable-by-variable Leave-One-Variable-Out (LOVO) jackknife sensitivity results across all $12$ candidate indicators for both Non-Negative PCA (NN-PCA) and unconstrained Standard PCA. Each row reports the Pearson linear correlation ($r$) and Spearman rank correlation ($\rho$) between the full $12$-indicator index and the $11$-indicator index re-estimated after omitting the specified variable..}
\end{tabular}
\end{table}

\end{document}